\documentclass[12pt,preprint]{aastex}
\usepackage{graphicx}

\slugcomment{Accepted for publication in ApJS}

\shorttitle{NLSy1 in barred galaxies}
\shortauthors{Ohta et al.}

\def\ltsima{$\; \buildrel < \over \sim \;$}
\def\lsim{\lower.5ex\hbox{\ltsima}}
\def\ltsim{\lower.5ex\hbox{\ltsima}}
\def\gtsima{$\; \buildrel > \over \sim \;$}
\def\gsim{\lower.5ex\hbox{\gtsima}}
\def\gtsim{\lower.5ex\hbox{\gtsima}}

\begin{document}

\title{A Bar Fuels a Super-Massive Black Hole?:  \\
Host Galaxies of Narrow-Line Seyfert 1 Galaxies
\footnote{Based on data collected at University of Hawaii 88inch telescope,
Canada France Hawaii Telescope, Subaru Telescope which is operated by 
the National Astronomical Observatory of Japan,
and Kitt Peak National Observatory 2.1m telescope, 
which is operated by the National Optical Astronomy Observatories
(NOAO), operated by AURA, Inc., under contract with the National Science 
Foundation.}}

\author{Kouji Ohta}
\affil{Department of Astronomy, Kyoto University,
Kyoto 606-8502, Japan}
\email{ohta@kusastro.kyoto-u.ac.jp}

\author{Kentaro Aoki}
\affil{Subaru Telescope, National Astronomical Observatory of Japan,
    650 North A'ohoku Place, Hilo, HI 96720}

\author{Toshihiro Kawaguchi\footnote{Present address:
Department of Physics and Mathematics, Aoyama Gakuin University,
Sagamihara, Kanagawa 229-8558, Japan}}
\affil{Optical and Infrared Division, 
National Astronomical Observatory of Japan, Mitaka, Tokyo 181-8588, Japan}

\and

\author{Gaku Kiuchi}
\affil{Department of Astronomy, Kyoto University,
Kyoto 606-8502, Japan}


\begin{abstract}
We present optical images of nearby 50 narrow-line Seyfert 1
galaxies (NLS1s) which cover all the NLS1s at $z<0.0666$ and 
$\delta \ge -25^{\circ}$ known at the time of 2001.
Among the 50 NLS1s, 40 images are newly obtained by our 
observations and 10 images are taken from archive data.
Motivated by the hypothesis that NLS1s are in an early phase 
of a super-massive black hole (BH) evolution, we present a study 
of NLS1 host galaxy morphology to examine  trigger mechanism(s)
of active galactic nuclei (AGNs) by seeing the early phase of AGN.
With these images, we made morphological classification 
by eye inspection and by quantitative method, and found 
a high bar frequency of the NLS1s in the optical band;
the bar frequency is $85 \pm 7$\% among  disk galaxies 
($64- 71$\% in total sample) which is more frequent than
that ($40-70$\% ) of broad-line Seyfert 1 galaxies (BLS1s)
and normal disk galaxies, though the significance is marginal. 
Our results confirm the claim by  Crenshaw et al. (2003) 
with a similar analysis for 19 NLS1s.
The frequency is comparable to that of HII/starburst galaxies.
We also examined the bar frequency against  width of the 
broad H$\beta$ emission line, Eddington ratio, and black 
hole mass, but no clear trend is seen.
Possible implications such as an evolutionary sequence 
from NLS1s to BLS1 are discussed briefly.
\end{abstract}


\section{INTRODUCTION}
Narrow-line Seyfert 1 galaxies (NLS1s) are a subclass of active
galactic nuclei (AGNs) which have the following characteristics 
(see Pogge 2000):
(1)They have relatively narrower permitted lines
(full width at half-maximum (FWHM) of H$\beta \le$ 2000\,km 
s$^{-1}$) than those of usual broad-line Seyfert 1 (BLS1s).
(2)Their X-ray spectra are significantly softer (photon index ($\Gamma$)
in a soft X-ray band is 1.5--5) than those of BLS1s 
($\Gamma \sim 2.1$) (Boller et al. 1996).
(3)They show rapid soft/hard X-ray variability (Leighly 1999).
(4)They often emit strong Fe II multiplets or higher ionization iron
lines, that are seen in Seyfert 1s but not seen in Seyfert 2s
(Osterbrock \& Pogge 1985). 

The most attractive and likely interpretation of the characteristics 
above is that NLS1s contain less massive black-holes (BHs) 
with a high accretion rate for a given luminosity, as described below. 
If the distance between the clouds emitting the broad-component of H$\beta$ 
and the central BH ($R_{BLR}$) is determined 
by dust-sublimation (Netzer \& Laor 1993) or 
by the intensity of the ionizing UV on the clouds 
(Baldwin et al. 1995; Korista et al. 1997), 
$R_{BLR}$ would scale with the luminosity of the central accretion disk 
($L$) as $R_{BLR} \propto L^{1/2}$ (e.g., Kaspi et al. 2000, 2005).
By assuming that the clouds in BLR are virialized 
(Peterson \& Wandel 1999, 2000) and 
that the luminosity $L$ is roughly in proportion to the accretion rate 
($\dot{M}$), 
the FWHM of the broad H$\beta$ emission-line is proportional to 
$M_{BH}^{1/2} \dot{M}^{-1/4}$ ($M_{BH}$ refers to a BH mass.).
Then, the relatively narrow H$\beta$ emission-lines can be
attributed to a smaller $M_{BH}$ and a higher $\dot{M}$.
This view  is also supported by results of 
reverberation mapping; smaller $M_{BH}$s and higher Eddington ratios 
for objects with narrower H$\beta$ width (Kaspi et al. 2000; Collin 
et al. 2002; Peterson et al. 2004).
The softness of the X-ray emission is also explained in the same hypothesis, 
because the maximum temperature of an accretion disk, 
which would determine the soft X-ray index, is predicted to scale as
 $M_{BH}^{-1/2} \dot{M}^{1/4}$ in the standard accretion disk model 
(Shakura \& Sunyaev 1973).
Similar (but not exactly the same) $M_{BH}$- and $\dot{M}$-dependencies 
of the temperature hold even for an accretion rate larger than
the critical one below which the standard model works 
(Mineshige et al. 2000; Kawaguchi 2003).
Since a smaller $M_{BH}$-system (i.e. smaller Schwartzschild-radius 
and thus more compact system)
would be fluctuating more rapidly than a larger $M_{BH}$ system,
NLS1s are thought to harbor smaller mass BHs (Hayashida 2000).

Less massive BHs with high accretion rates suggest
that BHs in NLS1s have not yet been fed enough to become massive ones,
and their BHs are now rapidly growing (Mathur 2000; Kawaguchi et al. 2004).
If NLS1s are indeed in an early phase
of BH evolution, 
they can be key objects for studying formation and 
evolution of AGNs.
Detailed studies on nearby NLS1s will enable us to reveal 
the formation mechanism(s) and process(es) of central BHs 
in low-redshift universe,
which of course helps the understanding of QSO formation and evolution
in high redshift universe.

A scenario that QSO/AGN activity is triggered
by galaxy-galaxy interaction and/or by bar structure has been proposed
(e.g., Simkin et al. 1980; Noguchi 1988; Shlosman et al. 1990).
In observational studies, although the excess of companion galaxies
for Seyfert galaxies was claimed (Dahari 1984; Keel et al. 1985),
recent studies do not support it (Schmitt 2001; Laurikainen \& Salo
1995).
The fraction of barred spiral galaxies among Seyfert galaxies is also
comparable to that of normal spiral galaxies (Heckman 1978; 
Simkin et al. 1980; Ho et al. 1997; Mulchaey \& Regan 1997; 
Hunt \& Malkan 1999).

If NLS1s are indeed in an early phase of AGN evolution,
we may be able to see such features (companions, bars) 
more clearly in NLS1s.
Motivated by this speculation, we started imaging 
observations of nearby NLS1s in optical band.
During we had been conducting this program, 
Crenshaw et al. (2003) studied morphology of 
91 Seyfert 1 galaxies (13 NLS1s and 78 BLS1s) 
at $z \leq0.035$, which were taken with HST WFPC2 through the F606W
filter by Malkan et al. (1998), and  additional 
six more NLS1s at $z \leq 0.084$ from archival WFPC2 data 
taken through F814W or F547M filters.
They found that among 84 disk galaxies (17 NLS1s and 67 BLS1s) 
the NLS1s tend to reside in barred galaxies more frequently 
than BLS1s; 11  out of 17 NLS1s  (65\%) have bars, 
while only 25\% (17/67) of BLS1s have bars.
When the sample is further limited to FWHM less than 1000 km s$^{-1}$,
100\% (4/4) of the NLS1s show the bar structure.
Our sample in the present study includes 50 NLS1s
taken from  literatures and from our own observations,
and it covers all the known NLS1s at $z \leq0.0666$ 
(and $\delta \ge -25^{\circ}$).

In this paper, we present results of optical imaging observations of
the NLS1 sample and examine the morphology 
of their host galaxies, especially the frequency of bar structure.
Sample selection is presented in the next section,
and the data collection (mostly through our own observations)
is described in \S 3.
Morphology classification is made in \S 4 and
resulting bar frequency and its comparisons with other
samples are discussed in \S 5. 
Possible implications of the results
are discussed in \S 6, followed by the summary of the paper in \S 7.
We adopt a cosmological parameter set of $H_0 = 70$ km s$^{-1}$,
$\Omega_{\rm M}=0.3$, and $\Omega_{\Lambda}=0.7$.

\section{SAMPLE}

Our sample consists of 50 NLS1s at $z<0.0666$ 
($cz \le 20000$ km s$^{-1}$)
and $\delta \ge -25^{\circ}$,
and covers virtually all the NLS1s known to date as of 2001. 
The redshift limit is introduced in order to resolve the bar
structure; $1"$ corresponds to less than $1.3$\,kpc which
should be compared with a typical length (semi-major axis) of a global bar
of a few to several kpc (e.g., Elmegreen and Elmegreen 1985; Ohta et al.
1990; Ervin 2005).
The sample size is 2.6 times larger than that of
NLS1s (19) studied by Crenshaw et al. (2003).
Most of the present sample were taken from
``a catalogue of quasars and active galactic nuclei, 10th edition''
(V\'eron-Cetty \& V\'eron 2001) in VizieR service 
(Ochsenbein et al. 2000).
However, among 54 NLS1s in the catalog, we removed 14 objects
by checking their original spectra shown in the literatures
and our own spectrum with a higher spectral resolution
($R\sim3000$) taken with the GoldCam attached to 
the KPNO 2.1m telescope (details of the observing set up 
is described by Aoki et al. 2005);
they show H$\beta$ emission lines broader than 2000 km s$^{-1}$
or show type-2 feature.
We also extracted NLS1s from Boller et al. (1996) and Xu et al.
(1999) with  the same criteria, but not listed in the catalogue by 
V\'eron-Cetty \& V\'eron (2001).
Again, by examining the original spectra in the literatures
we chose four NLS1s.
In addition to these, six NLS1s which satisfy the criteria
were picked up from the Bright Quasar Survey (Schmidt and Green 1983) 
by checking the FWHM of the broad H$\beta$ emission-line presented by
Boroson and Green (1992), if it is less than 2000 km s$^{-1}$.
The resulting sample is listed in Table\ref{tblsample} in 
order of the width (FWHM) of H$\beta$ emission line.
Although the sample is heterogeneous, it was the largest
sample of nearby NLS1s at that time.

For each object, we calculate an optical continuum
luminosity ($\lambda L_{5100}$), a black hole mass ($M_{\rm BH}$), and
an Eddington ratio, and the values are listed in Table 1.
The optical continuum luminosity is derived from the
$B$ magnitude given by V\'eron-Cetty \& V\'eron (2003)
and is corrected for the Galactic extinction by using
the NASA/IPAC Extragalactic Database (NED) (Schlegel et al. 1998).
The $k$-correction was done by assuming $f_{\nu} \propto \nu^{-0.44}$
(Vanden Berk et al. 2001).
The black hole mass is calculated by using the FWHM of H$\beta$
broad-emission line and the optical continuum luminosity 
following the method by Kaspi et al. (2000, 2005).
The Eddington ratio is calculated from the Eddington luminosity
($L_{\rm Edd}$) and the bolometric luminosity ($L_{\rm bol}$),
which is assumed to be 13 $\lambda L_{5100}$
(Elvis et al. 1994).
It should be noted that the black hole masses and thus
the Eddington ratios have a large uncertainty, because
the estimated continuum luminosity ($\lambda L_{5100}$) includes 
the light both from
the nucleus and the host galaxy.
It is possible that roughly up to 50\% of the $B$ magnitude may 
come from the host galaxies (Surace et al.  2001;
Bentz et al. 2006), which would be the dominant source of 
uncertainty in the estimations.
 
\section{IMAGING DATA SOURCES} \label{Obs}

Most of the imaging data of the sample were collected through our
observations with University of Hawaii 88 inch telescope 
during the period from Apr 2003 to May 2005.
The imagers used were the OPTIC (Tonry et al. 2004) and
the Tek 2k camera.
The field of view and pixel size were $\sim 5^{\prime}$ and
0.$^{\prime\prime}$138 (or 0.$^{\prime\prime}$276 in
the 2 binning mode), respectively for OPTIC, and
$\sim 7^{\prime}$ and 0.$^{\prime\prime}$22, respectively
for Tek 2k.
We adopted $I_{\rm C}$-band to trace stellar continuum.
A typical exposure time of each frame was 2$-$5 minutes
to avoid saturation of a bright nucleus.
A total exposure time for each object was 10--30 minutes 
for most of them.
Seeing sizes during observations were 0.$^{\prime\prime}5$
 to 1.$^{\prime\prime}$4.
The weather condition was not photometric for most of the
observing runs.

We also observed a part of the sample in $g'$-band in March 2003 
with the MegaCam attached to Canada France Hawaii Telescope (CFHT) 
in Queue mode as a Snapshot program. 
The field of view was $\sim 1^{\circ}$ and the pixel size was
0.$^{\prime\prime}$185.
Each exposure time was 3 minutes, 
and a total exposure time for each object was 21--27 minutes.
The seeing size was around 0.$^{\prime\prime}$8 to 
1$^{\prime\prime}$4.
The weather condition was partly photometric.
We took an additional image with FOCAS (Kashikawa et al. 2002)
on the Subaru telescope (Iye et al. 2004) in $I_{\rm C}$
band in Nov. 2004.
Totally, we obtained 40 images out of the 50 targets.
The journal of the observations is shown in Table \ref{tblrun}.

The imaging data were reduced  with usual manner;
after subtracting bias, a flat fielding was applied.
Since the OPTIC moves charges at pixels during an exposure, 
a flat field frame cannot be made with usual manner.
Thus we made the flat frame by using `conflat' program developed 
by Tonry, which makes a flat frame by weighting net exposure times
and efficiencies of pixels where the charge was staying.

Figure 1 shows montages of all the NLS1s in our sample
in the order of H$\beta$ emission-line width from smaller FWHM
to larger FWHM (same as the order in Table 1).
For the 10 remainders  we took images from
HST archive, Sloan Digital Sky Survey (SDSS) 
Data Release 5 (DR5), and digitized Palomar Observatory Sky Survey (POSS),
though quality of some of the POSS data are poor.

Since the weather conditions were not good, it is impossible
to make photometric calibration for these targets with our data.
Nevertheless, in order to assess the depth of the imaging data
roughly, we made the calibration as follows.
For about 30\% of the sample, SDSS  photometric data are available.
For these objects, we derived $I$ magnitudes 
from SDSS $i$ magnitudes and $i-z$ colors using a recently obtained
magnitude transfer equation by Lupton (2005)\footnote{
www.sdss.org/dr4/algorithms/sdssUBVRITransform.html}
for UH88 data, $g'$ magnitudes from $g$ magnitudes for
CFHT data, and $R$ magnitude from $r$ magnitude and $r-i$ color
for POSS data.
For other objects, we derived $I$- (UH88 and Subaru data) 
and $g'$- (CFHT data), and $R$- (POSS data) magnitudes from
the cataloged $B$ magnitudes by using the transformation
equations (Lupton 2005) with use of average colors of
$g-r$, $g-i$, $i-z$, and $r-i$  calculated from
the colors  of the above NLS1s having  SDSS data
($<g-r> = 0.43 \pm 0.36$ mag, $<g-i>=0.75\pm0.17$ mag, 
$<i-z>=0.18\pm0.09$ mag, and $<r-i>=0.33\pm0.06$ mag).
Since the standard deviations of the average colors shown
are not very small, it should be kept in mind that
this estimation may have uncertainty of up to 1.0 mag,
and only gives a rough idea on the surface brightness
achieved by each imaging data.
As for the data taken with HST (F606 and F814) and the 
SDSS data ($i$-band), 
the photometric calibration was done by using its FITS header
and the total magnitude in the catalog, respectively.
The radial surface brightness distributions derived
by using `ellipse' task in IRAF are shown in Figure 1.


\section{MORPHOLOGY OF NARROW-LINE SEYFERT 1s}

\subsection{Morphology by Eye Inspection}
We assigned a morphological type for each object 
by eye inspection based on the images (Figure 1).
When the global bar structure is clearly seen, we assign SB.
If the bar seems to be weak or oval, SAB is assigned.
Hubble sequences of 0/a ($T=0$), a ($T=1$), ab ($T=2$), 
b ($T=3$), bc ($T=4$), and c ($T=5$) are also assigned;
a typical accuracy would be $\pm (1-2)$ in $T$.
When an interaction/merge or peculiar feature is seen,
we assign `int' or `P', respectively.
Although `?' mark is labeled 
when we are not perfectly convinced of the classification,
we include them for the following statistics unless otherwise
mentioned.

We made eye  inspection again  several months after 
the first assignment of morphology without looking at the 
previous assignment.
The results are almost consistent.
For 95\% of the sample, the morphology agrees with  each other
within $\Delta T = \pm 1$.
Three  SB galaxies are classified as SAB, one SAB
is classified as SB, and two SAs as SABs;
consequently the  changes in the total number of SB and SAB are
$-2$ and $+4$, respectively.
The uncertainty in the classification does not affect 
so much for SBs and SBs+SABs, because most of the NLS1s are classified as SB
or SAB as described below.
The resulting assignment of morphology is shown in Table 
\ref{tblsample} together with the data sources of the eye
inspection.

Morphology in parenthesis in Table \ref{tblsample} is  that
shown in Crenshaw et al. (2003).
These classifications were also made by eye inspection.
For four objects, morphology assignment is different from those by us;
we assign SAB for three objects classified as SA by Crenshaw et al. (2003)
because there seems to be a weak bar-like structure, while
for one case we cannot recognize the strong bar structure 
and hence assign SAB instead of SB.

We compared our classification with those in the Third Reference
Catalogue of Bright Galaxies (RC3; de Vaucouleurs, et al. 1991).
Since our sample contains galaxies with rather large redshifts,
most of the galaxies are not listed in RC3. 
14 objects are classified in RC3, but among them eight objects are 
classified with `?' (e.g., `S?'), and thus it is not meaningful
to make comparison.
Among the remaining six objects which we assigned SB, 
four objects are classified as SB, one object (NGC 4051) as
SAB, and one object (Mrk 1044) as peculiar in RC3.

\subsection{Quantitative Classification of Morphology}

We also tried to make a quantitative classification of the bar structure.
There are several ways to detect a bar; Fourier analysis of azimuthal
profiles (e.g., Buta and Block 2001), ellipse fitting  of isophotes
(e.g., Mulchaey et al. 1997), axial ratio and twist of isophotes
(Abraham et al. 1999), and variations of these methods.
Here we adopt the method with use of radial variation of
ellipticity and position angle of isophotes of a galaxy,
because now it seems to be widely used and is rather robust for a galaxy
with a relatively small angular extent.
Following  Jogee et al. (2004), we employ the following criteria
to identify the bar:
(1)the ellipticity should have global maximum value ($e_{\rm max}$)
of larger than 0.25 and the value should be larger than that of
the outer disk,
(2) the position angle (PA) in the bar region should not change 
larger than $\pm 20^{\circ}$, and
(3)the ellipticity should drop by $\ge 0.1$ ($\Delta e \ge 0.1$)
and the PA usually changes by $\ge 10^{\circ}$
at around the bar end.

We used `ellipse' task in IRAF and examined the radial
variation of the ellipticity and the position angle.
When a companion or a star is located close to the target object,
we masked it out.
The resulting profiles are shown in lower panels of right side of 
Figure 1; solid circles and solid squares refer to the ellipticity 
with the left side ordinate and the position angle with right side
ordinate, respectively.
A horizontal dotted line, dashed line, and dot-dashed line
show $e=0.25$, global maximum value of ellipticity $e_{\rm max}$,
and $e_{\rm max}-0.1$, respectively.
The latter two lines are shown only for the objects classified 
as SB or SAB.

Results of the quantitative identification are shown in column 10 of
Table 1; 
when the criteria are satisfied, we assign `SB', while
the criteria are not satisfied, `no' is assigned, 
though it may not appropriate for Mrk 335, in which
the nucleus dominates the light distribution and we may not trace
its host.
In two cases (Zw493.004 and NPM1G $-$14.0512),
$\Delta e \sim 0.09$ were obtained with $e_{\rm max} > 0.25$.
We assign `SAB' for these objects, though usual quantitative
classification does not assign `SAB'.
Since Zw493.004 and NPM1G $-$14.0512 show a clear bar structure,  
we judge it is reasonable to assign `SAB'.
For IZw1, we obtained $e_{\rm max} > 0.25$ and the constant 
position angle, but we assign `no', because $e_{\rm max} > 0.25$ 
is seen only in one data point.
PG 1535+547 and WAS 61 show large $e_{\rm max}$s of $0.4 - 0.5$ and 
$\Delta e>0.1$.
Although these may be edge-on galaxies, we assign `SB' following
the criteria.
For some of the targets we failed to make the analysis
due to the presence of a strong interaction or merger
(RX J0140.1+1129, Mrk 507, Mrk 739E, and KAZ 163), or 
due to the presence of a clear feature such as spiral arm of which 
isophotes deviate from an ellipse significantly (NGC 4051).
In these cases, we assign `$-$' in Table 1 and do not show
the radial profiles of ellipticities and position angles.

As seen from Table 1, most (21/24, 88\%) of the SB galaxies 
assigned by eye inspection are identified with SBs in the quantitative
classification.
Differences with respect to the classification by eye inspection 
are as follows:
Three  SAs in the eye inspection are identified with three SBs
in the quantitative classification,
eight SABs with six SBs and two unbarred galaxies (i.e., `no'),
and three SBs with two SABs and one `$-$'.
Hence  the total numbers of SABs and SBs in eye inspection,
respectively, are six more and six less compared with those 
in the quantitative classification.
But the total number of SBs plus SABs does not change.

We also compared the results with the classification in RC3.
Among the six objects, which are classified without `?' in RC3, 
three  objects we assigned SB are classified as SB,
one  object (Zw493.004) we assigned SAB is classified as SB,
one object (Mrk 1044) we assigned SB is classified as P,
and one object (NGC 4051) which we could not make quantitative
classification is classified as SAB.

There seems to be advantages and disadvantages to use
either of the classification method.
Morphology assignment is different from each other
for a part of the sample, which gives different resulting bar frequencies.
Therefore,  we adopted both of the classification methods and 
discuss the bar frequencies  using the two classifications.

The quantitative classification can distinguish  the presence or
the absence of the bar.
It cannot, however,  give us information about the disk or elliptical
for unbarred galaxies.
Thus we supplementary examined the images of the unbarred
galaxies and their radial surface brightness distribution 
to see whether  it shows the exponential-law or $r^{1/4}$-law distribution.
PG 1244+026, RX J1531.6+2019, IRAS 15091$-$2107, and KAZ 320
show the $r^{1/4}$-law dominated profiles, though IRAS 15091$-$2107 
may show the presence of a slight disk component.
These galaxies are classified as E/S0s in the eye inspection.
RX J1032.7+3913 and RX J1402.5+2159, which are classified as E/S0?
in the eye inspection, also show the $r^{1/4}$-law like profiles.
RX J1618.1+3619, Mrk 1239, IZw1, RX J17450+4802, and PG0923+129
show disk dominated features and they are all classified as disk galaxies
in the eye inspection.
Remaining two galaxies, PG 1448+273 and Mrk 335 are interacting or
peculiar galaxies.
Thus for the E/S0 classification we adopt the eye inspection 
in  the quantitative classification.
When we take a disk-galaxy sample in the following statistics, 
we exclude all the galaxies classified as `$-$' as well as E/S0s, 
and interacting or peculiar galaxies among unbarred galaxies.

\section{FREQUENCY OF BARRED GALAXIES}

\subsection{Bars in NLS1s}

As seen in Figure 1, the frequency of barred galaxies is fairly high.
Among 50 NLS1s for which we assigned morphological type by eye inspection,
24 NLS1s show clear bar structure (SB) and 8 NLS1s show weak or 
oval structure (SAB),
which leads to the bar frequency of $48 \pm 7$\% (24/50) and
$64 \pm 7$\%\footnote{The 
uncertainty for the frequency is estimated from 
$\sigma^2 = (1-f) f / N$, where $f$ is an observed frequency of 
bars in the sample and $N$ is its sample size.}
(32/50) for SB and SB+SAB, respectively (Table 3).
(If we exclude galaxies classified with `?', they are
$55\pm8$\% (23/42) and $69\pm7$\% (29/42), respectively.)
The frequencies are comparable in the quantitative classification;
the SB frequency is $60\pm7$\% (30/50) and the SB+SAB frequency is
$64\pm7$\% (32/50).
If we exclude the galaxies which cannot be classified (i.e., `$-$'),
the SB and SB+SAB frequencies go up to $67 \pm 7$\% (30/45) 
and $71 \pm 7$\% (32/45),  respectively (Table 4).
Hereafter statistics derived with the quantitative classifications
always exclude the five  objects classified as `$-$'.
 
Following  Crenshaw et al. (2003), we see the frequencies
among disk galaxies by excluding  E and E/S0 galaxies, 
interaction or peculiar galaxies, and no morphological assignment
(`$-$' in Table 1) as well from the sample.
The resulting bar frequency  amounts up to $63 \pm 8$\% (24/38)
and $84\pm6$\% (32/38) for SB and SB+SAB, respectively in the
classification by eye inspection (Table 3).
(If we exclude galaxies classified with `?', they are $72\pm8$\%(23/32)
and $91\pm5$\% (29/32), respectively.)
The SB and SB+SAB frequencies are $81\pm6$\% (30/37) and $86\pm6$\% (32/37),
respectively in the quantitative classification (Table 4).
If we take NLS1s with H$\beta$ FWHM less than 1000 km s$^{-1}$,
the frequency (SB+SAB) goes up to $90 \pm 9$\% (9/10) in both classifications,
confirming 
the results  by Crenshaw et al. (2003).
Below we adopt the value of $85\pm7$\% as the bar frequency of
the NLS1s among disk galaxies, considering the frequencies in
the eye inspection and the quantitative classification.
It should be kept in mind, however,  that the uncertainty  is a
statistical one and it does not include the uncertainty of
the classification;
if we miss-classified one galaxy to be a barred galaxy,
then the frequency changes about 3\%.
Since the difference in the total number of SBs is two between the
two eye inspections and we have two SABs in the quantitative
classification, this number would give an estimation
of the uncertainty of the classification.

We  see the bar frequencies against distance of galaxy
to examine whether our classification shows some incompleteness
or bias.
We divide the sample into two subsamples with a comparable
size number; nearby subsample ($z<0.041$, 24 galaxies) 
and distant subsample ($z>0.041$, 26 galaxies).
In the eye inspection classification, the bar frequency (SB+SAB)
in the nearby subsample is 75$\pm9$\% (18/24), while
it is 54$\pm10$\% (14/26) in the distant subsample.
In the quantitative classification, the fractions in
the nearby and distant subsamples are 77$\pm9$\% (17/22) and 
65$\pm10$\% (15/23), respectively.
The cause for the lower frequency in the distant sample
is the presence of E/S0 galaxies and of 
interacting or peculiar galaxies.
When we calculate the bar frequency among disk galaxies,
the bar frequencies are $86\pm8$\%  (18/21) and $82\pm9$\% (14/17)
for the nearby and distant subsamples, respectively in
the eye inspection classification.
In the quantitative classification, the frequencies are 
$85\pm8$\% (17/20) and $88\pm8$\% (15/17) for the nearby 
and distant subsamples, respectively.
Thus in both classifications, the frequencies among disk
galaxies agree with each other and they do not change 
against the distance.

Next we  examine the dependency of the frequencies against 
H$\beta$ FWHM.
The bar (SB+SAB) frequency among the total sample
seems to decrease with increasing H$\beta$ FWHM in
 the eye inspection classification;
$82 \pm 12$\% (9/11), $63 \pm 11$\% (12/19), and $55 \pm 11$\% (11/20)
for FWHM of $<1000$ km s$^{-1}$, 1000 -- 1500 km s$^{-1}$,
and 1500 -- 2000 km s$^{-1}$, respectively (Table 3).
In the quantitative classification,  although the frequency is
high among galaxies with H$\beta$ FWHM less than 1000 km s$^{-1}$,
no such clear trend is seen;
$82\pm12$\% (9/11), $56\pm12$\% (10/18), and $81\pm10$\% (13/16),
for FWHM of $<1000$ km s$^{-1}$, 1000 -- 1500 km s$^{-1}$,
and 1500 -- 2000 km s$^{-1}$, respectively (Table 4).
If we take the frequencies among disk galaxies,
the frequencies for the three cases are 
$90 \pm 9$\% (9/10), $86 \pm 9$\% (12/14), and $79 \pm 11$\% (11/14)
for FWHM of $<1000$ km s$^{-1}$, 1000 -- 1500 km s$^{-1}$,
and 1500 -- 2000 km s$^{-1}$, respectively for the eye inspection
 classification (Table 3), and 
$90 \pm 9$\% (9/10), $77 \pm 12$\% (10/13), and $93\pm 7$\% (13/14)
for FWHM of $<1000$ km s$^{-1}$, 1000 -- 1500 km s$^{-1}$,
and 1500 -- 2000 km s$^{-1}$, respectively for the quantitative
classification (Table 4).
Thus no clear trend is seen among disk galaxies.

We here recall the original motivation of this imaging program.
NLS1s likely have smaller $M_{BH}$ and higher $\dot{M}/M_{BH}$ ratios.
Thus, the elapsed time since central BHs begun to grow by accretion 
events ($\ltsim M_{BH} / \dot{M}$ assuming 
quasi-steady accretion) is likely to be shorter for NLS1s 
on average than for BLS1s.
Such ``young AGNs'' may  still keep (while ``older AGNs'' could have lost)
traces of unknown process(es) that triggered an major accretion event
towards a central BH.
For this purpose, sorting the sample NLS1s by the Eddington ratio
(roughly scales with $\dot{M}/M_{BH}$ ratio) would be more 
physically meaningful than sorting by line width examined above.
We divide the NLS1s into three subsamples,
according to their Eddington ratios.
The resulting bar frequencies (SB+SAB) for subsamples with 
$L_{\rm bol}/L_{\rm Edd} = 10^{-0.39}-10^{-0.1}$, $10^{-0.1} -
10^{0.3}$, and $10^{0.3} - 10^{1.0}$
are $55 \pm 11$\% (11/20), $63 \pm 11$\% (12/19), and
$82 \pm 12$\% (9/11), respectively in the eye inspection
classification (Table 3).
There seems to be a rough correlation between the bar frequency and
Eddington ratio.
In the quantitative classification, however, no trend can be seen;
the bar frequencies are $75 \pm 11$\% (12/16), 
$67 \pm 11$\% (12/18), and $73 \pm 13$\% (8/11)
for $L_{\rm bol}/L_{\rm Edd} = 10^{-0.39} - 10^{-0.1}$, $10^{-0.1} -
10^{0.3}$, and $10^{0.3} - 10^{1.0}$, respectively (Table 4).
If we take the frequencies among the disk galaxies,
the bar frequencies are high and 
the trend is not clear;
$79 \pm 11$\% (11/14), $80 \pm 10$\% (12/15), and 100\% (9/9)
for $L_{\rm bol}/L_{\rm Edd} = 10^{-0.39} - 10^{-0.1}$, 
$10^{-0.1} - 10^{0.3}$, and $10^{0.3} - 10^{1.0}$,
respectively in the eye inspection classification (Table 3).
The situation is the same when we take the quantitative classification;
the frequencies are $92 \pm 7$\% (12/13), $80\pm10$\% (12/15),
and $89\pm10$\% (8/9) for 
$L_{\rm bol}/L_{\rm Edd} = 10^{-0.39} - 10^{-0.1}$, 
$10^{-0.1} - 10^{0.3}$, and $10^{0.3} - 10^{1.0}$,
respectively (Table 4).
Thus no significant trend is seen.
 
We also examine the bar frequency against $M_{\rm BH}$.
The frequencies of SB + SAB are
$62 \pm 13$\% (8/13), $68 \pm 10$ \% (15/22), and $60 \pm 13$ \% (9/15)
for subsamples of $M_{\rm BH} = 10^{5.35} - 10^{6.5} M_{\odot}$,
$10^{6.5}-10^{7.0} M_{\odot}$, and $10^{7.0} - 10^{7.5} M_{\odot}$,
respectively in the classification by eye inspection (Table 3).
The frequencies in the quantitative classification range
from 67 \% to 75 \% (Table 4) and no trend is seen.
If we take statistics among disk galaxies,
the bar frequencies are
$73 \pm 13$\% (8/11), $88 \pm 8$\% (15/17), and $90 \pm 9$\% (9/10),
for $M_{\rm BH} = 10^{5.35} - 10^{6.5} M_{\odot}$,
$10^{6.5}-10^{7.0} M_{\odot}$, and $10^{7.0} - 10^{7.5} M_{\odot}$,
respectively in the eye inspection classification (Table 3).
The frequencies are $80 \pm 13$\% (8/10), $88 \pm 8$\% (15/17),
and $90 \pm 9$\% (9/10) for $M_{\rm BH} = 10^{5.35} - 10^{6.5} M_{\odot}$,
$10^{6.5}-10^{7.0} M_{\odot}$, and $10^{7.0} - 10^{7.5} M_{\odot}$,
respectively in the quantitative classification (Table 4).
Although in the eye inspection classification the frequencies might
show a slight increase with increasing black hole mass, 
no trend is also acceptable within the errors.
Hence there is no clear trend against $M_{BH}$.

\subsection{Comparisons of Bar Frequencies} \label{discussion1}

Since we collected all the known NLS1s,  it is a kind of heterogeneous sample 
(though we set the limit on their distance, $z$), and thus 
it is not obvious what kind of control sample we should take to compare.
Nevertheless, as described below, many `control' samples point
to the canonical value of bar frequency.
First, we confront the bar frequency of NLS1s with that of nearby 
galaxies.
Classically, for instance, de Vaucouleurs (1963) made a classification of 
$\sim$1500 bright galaxies in the local universe and
65\% of disk galaxies were classified
as barred galaxies (SB (37\%) and SAB(28\%)).
More recently, Hunt \&  Malkan (1999) examined the frequency of
barred galaxies by sampling galaxies selected from
the Extended 12 $\mu$m Galaxy Sample (Rush et al. 1993),
and found 68--69\% (with a typical statistical uncertainty of 
$\pm 2$\%) of non-active normal galaxies (i.e., excluding
Seyferts, LINERs, HII/starburst galaxies) are classified 
as strong bar (SB) or weak bar (SAB), based on RC3.
Laurikainen et al. (2004) also examined the bar frequency  of
non-active galaxies for a sample ($M_{\rm B} < 12$ mag, diameter
$< 6\farcm5$,  inclination $<60^{\circ}$, etc
from RC3) and found again 57--69\% (with a typical uncertainty of
$\pm 5$\%) are SB or SAB.
In a magnitude limited sample ($B_T \le 12.5$ mag and $\delta >
0^{\circ}$),
Ho et al. (1997) showed that 40--80 \% of disk galaxies
are classified as SB or SAB based on RC3.

The frequency of SB and SAB varies with Hubble type;
the percentage increases in later type  (Ho et al. 1997).
Since almost all the NLS1s in our sample is  earlier than Sc,
considering galaxies earlier than Sc is reasonable
for comparison.
For such galaxies, Ho et al. (1997) found the bar frequency 
of 40-60\%.
It is worth noting that photometric properties of bars in
early-type are quite different from those in late-type,
and its boundary is around SBbc; e.g., surface brightness distribution
of the early-type bar is flat with a sharp cutoff at the bar end
while that of the late-type is steep without a sharp cutoff
  (Elmegreen \& Elmegreen 1985; Ohta 1996).
Hence the origin of bars may be different
between bars in early- and in late-type disk galaxies.

In summary, all the samples of which criteria are different each other
show the bar frequency of 40 to 70\% for normal disk galaxies.
Therefore, the bar fraction (SB+SAB) of $85 \pm 7$\%
among disk NLS1s in both the eye inspection  and the quantitative 
classifications is high as compared with that of normal disk galaxies.
If we adopt the average bar frequency for normal disk galaxies
as 70\% (60\%), then the significance of the excess would be
2.1$\sigma$ (3.6$\sigma$).
However it should be noted that this significance level does not include
the miss-classification in our morphology assignment as well as
that in other studies. 
For instance, if we miss-classify one object in our sample, 
the significance level reduces (or increases) about 0.5$\sigma$.

Next, we confront the value with that for Seyfert galaxies
(and LINERs).
According to Hunt \& Malkan (1997),
the bar frequency (SB and SAB) depends on a sample chosen,
but it ranges 56\% to 70\% with a statistical uncertainty
of $\sim10$\% for each sample.
They claim that the frequency  is comparable to
that of non-active galaxies and that no significant excess of
the bar frequency can be seen.
Laurikainen et al. (2004) also examined the bar frequency and
found  56--62\% (with an uncertainty of $\pm 7\%$ for each sample)
for Seyferts and LINERs.
Again they claimed no significant bar excess against non-active
galaxies.
The same conclusion was obtained by Ho et al. (1997).
Therefore, the frequency of bar seen in NLS1s ($85\pm7$\% in disk galaxies) 
is high  as compared with Seyferts/LINERs
with the similar significance level for normal disk galaxies.

It is also interesting to see the bar frequency among HII/starburst
galaxies.
Hunt \& Malkan (1999) found that the bar frequency (SB+SAB) of
 HII/starburst galaxies is high (82--85\%) (with an uncertainty of
$\pm$(6--7)\% for each sample).
The high  frequency ($78\pm9$\%) is also found in a study by
Laurikainen et al. (2004).
These high bar frequencies may occur  by inclusion of
late type galaxies.
However distribution of the frequency against the Hubble type
does not show such trend clearly (Hunt \& Malkan 1999),
and they are $57\pm6$\% and $64\pm5$\% among early (S0/a -- Sbc) and 
late (Sc -- Sm) type galaxies, respectively (Ho et al. 1997).
Although the bar frequency by Ho et al. (1997) is not so high, they are
comparable to each other.
The high occurrence of bar in HII/starburst galaxies seems 
to be a general trend, and the frequency seems to be comparable 
to that of NLS1s.

We note that the bar frequencies described above 
are all based on optical images.
Since the optical light is affected by dust extinction,
NIR is more suitable to recognize stellar bars.
In fact,  Mulchaey \& Regan (1997) made $K$-band imaging
and found the bar frequency  of $\sim 70$\% for normal galaxies.
Seigar \& James (1998) even found that almost 90\% of their sample
spirals (45) show bar structure in $J$ and $K$-band.
Eskridge et al. (2000)  made $H$-band imaging observations
for 186 spiral galaxies drawn from  the Ohio State University
Bright Spiral Galaxy Survey and found 56\% are strong bar and 
16\% are weak bar, totaling 72\% of the bar frequency.
Although Knapen et al. (2000) found the bar frequency of 59\%
in spite of the use of NIR data,
NIR data tend to give a higher bar frequency  for normal galaxies 
than optical data.
Since our observations were made in optical bands,
it is not adequate to compare our results with these NIR values.
However, we need to pay attention that $I_{\rm C}$-band and 
$g'$-band we used cover slightly longer wavelength than those 
described above for the optical bar frequency (mostly $B$- or $V$-bands).
Thus we cannot completely rule out a possibility that 
the high bar frequency found in this study 
is affected by the band effect,
although it is unknown how much degree the band effect
is serious in $I_{\rm C}$-band and $g'$-band.

\subsection{Frequency of Ring and Companion/Interaction}

We briefly mention the frequency of ring structure among NLS1s.
Hunt \&  Malkan (1999) found a high ring frequency,
particularly the frequency of an outer ring ($\sim 10$ kpc in radius)
among Seyfert galaxies.
The frequency of outer rings, including pseudo outer rings, is
19--41\% (depending on subsamples with a statistical
uncertainty of $\sim 10$\% for each subsample) which contrasts to
4--11\% for normal or HII/starburst galaxies.
If our NLS1s also have the same high outer ring frequency,
10 to 20 galaxies with the outer ring are expected.
However, in our sample, only  one  NLS1 shows a clear outer ring
(PG 0923+129), and we do not find many NLS1s with the outer ring.
The surface brightness of the outer ring in PG 0923+129 is
$\mu_g \sim 22.5$ mag arcsec$^{-2}$, which 
seems not be too faint to detect the similar outer ring 
in other galaxies in our sample, 
because we reached the fainter surface brightness
for most of the sample galaxies. 
Thus, it is possible that the frequency of outer rings associated 
with NLS1s is likely to be much smaller than that of BLS1s, 
provided that the surface brightness of the outer ring of PG 0923+129 
is typical.
However, whether the surface brightness of the outer ring is
almost universal or not is unclear and our surface brightness 
estimation has a large uncertainty, hence  further examination with 
deeper images is desirable to be definitive.

The  number of interacting galaxies and mergers (including
peculiar morphology) is  not large in our sample;
among 50 NLS1s four to eight NLS1s (8--16\%) show such features.
This is a rather small number if we compare it with the
results for Seyfert galaxies e.g., 20--30 \% by Schmitt (2001).
This contrasts the result by 
Krongold et al. (2001) who
studied environment of NLS1s and BLS1s using
the Digitized Sky Survey data, and found no systematic difference 
in the environments between them.
However, the quantitative comparison is not straightforward,
since we do not have  redshift information of close companions 
of the sample galaxies. 

It is worth noting that the fraction of
interacting galaxies and mergers (including peculiar) tends to be larger 
among NLS1s with larger FWHM of H$\beta$ emission lines.
For the NLS1s with FWHM smaller than 1000 km s$^{-1}$, 
no such galaxies are found, i.e., 0\% (0/11).
The frequency is $5\pm5$\% (1/19) for NLS1s with FWHM
between 1000 km s$^{-1}$ and 1500 km s$^{-1}$.
For the NLS1s with FWHM larger than 1500 km s$^{-1}$
(and less than 2000 km s$^{-1}$ by definition),
it is $35\pm11$\% (7/20).
If we discard the interacting galaxies classified  as `?',
the frequency is $20\pm9$\% (4/20).
Since it is unlikely that the tidal force affects the
line width of the broad line region, the cause for this
tendency is not clear.

\section{IMPLICATIONS OF THE DISTINCT HOST MORPHOLOGY}
Although the excess of the bar frequency in NLS1s as compared with
those of normal disk galaxies as well as of BLS1s is
marginal, we briefly discuss possible
implications in case that the morphology distribution 
of NLS1s is really different from those of normal disk 
galaxies and BLS1s.

\subsection{Are Narrow-line Seyfert 1s Broad-line Seyfert 1s
with Different Viewing Angles?}

The difference in the morphology distribution of host galaxies
between NLS1s and BLS1s gives an important implication to the 
nature of NLS1s.
Although many observational properties point to that the NLS1s
are different population from BLS1s, there has been
the claim that the NLS1s are BLS1s with different viewing
angles:
face-on view of a disk-like BLR (Puchnarewicz et al. 1992;
Boller et al. 1996; Taniguchi et al. 1999)
or partial obscuration (angles somewhere between Seyfert 1s and 2s).
Disks of host galaxies and accretion disks are not necessarily 
aligned (Schmitt et al. 2002). 
Thus if NLS1s are  objects with such specific viewing angles,
there should not be any difference in 
morphology of host galaxies between NLS1s and BLS1s.
Therefore, the difference in the host morphology
between NLS1s and BLS1s (more frequent bars and
rare outer rings in NLS1s) disproves the simple hypotheses 
above involving the viewing angle of the central engine,
though the difference is not so significant.

\subsection{Evolutionary Sequence of Seyfert Galaxies?}

The difference of  morphology frequency of the host galaxies 
of NLS1s against BLS1s (more frequent bars and rare outer rings)
and the similarity of it to HII/starburst galaxies
(comparable bar frequency) suggest the following evolutionary 
sequence scenario of these populations.
In actively star-forming galaxies with bar structure,
the gas accretion towards the central BH (i.e., ignition of 
AGN activity) is induced by bar by transforming gas angular
momentum outside of the bar region
(e.g., Simkin et al. 1980; Noguchi 1988; Shlosman et al. 1990).
Since a gas accretion rate onto a central BH is likely 
determined by external reasons
such as bar strength or amount of gas 
rather than internal reasons such as radiation pressure 
or gas outflow from the vicinity of the BHs (Collin \& Kawaguchi
2004),
$\dot{M}$ does not necessarily follow the increase of
BH mass with time.
Accordingly, if we assume that $\dot{M}$ is nearly constant 
during the lifetime of an AGN activity, BHs will be fed 
via super-Eddington accretion rates
(i.e. NLS1 phase) in the beginning, and 
via sub-Eddington rates (BLS1 phase) later on 
due to the increase of BH mass.
If the BLS1 phase indeed follows the NLS1 phase
(e.g., Mathur 2000; Wandel 2002; Kawaguchi et al. 2004), the outer ring 
could be expected  to form due to the gas angular momentum
transfer to the outer region of the galaxy as an reaction of 
the gas accretion into the central region of the galaxy
(e.g., Hunt \& Malkan 1999).

\subsection{Bar Destruction?}

If a NLS1 indeed includes a young growing massive BH
and will evolve to a BLS1, the difference of the bar
frequency implies that bars of NLS1s should be  dissolved 
during the transition period from the NLS1 phase to the BLS1 phase.
(Of course, some fraction of host galaxies of BLS1 
show the bar structure, thus not all the bars of NLS1s should be
dissolved.)
One possibility is that a bar is destroyed by  central
mass concentration (CMC; 
e.g., Hasan \& Norman 1990);
a rapidly growing massive BH located at the center of a galaxy
could be a CMC and may dissolve the bar structure.
Recent numerical studies show that a few to several percent of 
a disk mass is necessary to dissolve the bar
(Shen \& Sellwood 2004; Athanassoula et al. 2005).
Although the central BH in a NLS1 seems to be slightly 
less massive ($\ltsim 10^7 M_{\odot}$) than this threshold,
the mass could increase up to $\sim 10^8 M_{\odot}$ or more
during the transition phase from NLS1 to BLS1.
Furthermore, if we regard the CMC as a central BH together with
the gas fallen into the central region of the galaxy,
the total (BH plus gas) mass would exceed the threshold.
Therefore, it is expected to be possible that
the bar destruction by central BH (and gas in the 
central region as well) can occur in terms of the threshold mass. 

The time-scale for the bar destruction is also an important clue.
The time-scale of the bar destruction is likely to be a few Gyr 
(Shen \& Sellwood 2004; Athanassoula et al. 2005),
which is about 10 times galactic rotation 
in a disk region.
Meanwhile, the duration (lifetime) of the NLS1 phase
is estimated to be  10--30\,Myr  based on the relative fraction 
of NLS1s among type 1 AGNs (Grupe 1996; Kawaguchi et al. 2004).
It can also be estimated from the $e$-folding time-scale
(Salpeter time-scale) for the Eddington-limited accretion.
A BH mass increases by a factor of $e^3$ during the period
of 120 Myr ($\sim 3$ times the Salpeter time-scale).
Since the FWHM is in proportion to 
$M_{\rm BH}^{1/2} \dot{M}^{-1/4}$ and thus to
$M_{\rm BH}^{1/4}$ in this case, 
it results in a factor of 2.1 increase in FWHM,
which makes most NLS1s evolve to BLS1s.
Thus, 120 Myr is another estimation for the lifetime of NLS1.
For cases with constant $\dot{M}$, on the other hand,
a factor of 2 increase in the FWHM requires a factor of
4 increase in $M_{\rm BH}$.
Since many of NLS1s show super-Eddington accretion rates
(Kawaguchi 2003; Collin \& Kawaguchi 2004), the time-scale
can be as small as 3 times Salpeter time-scale or even smaller.
If we accept super--Eddington accretion in NLS1s,
a significant increase of $M_{\rm BH}$ is possible during 30Myr 
(Kawaguchi et al. 2004).
If the NLS1 phase appears episodic, the durations of the NLS1 phase 
discussed above 
must be the sum of a number of shorter episodic phases.
In any cases, the duration of the NLS1 phase is considered to be
too short to dissolve  the bar.
Alternatively, the bar structure could be dissolved  gradually
during the BLS1 phase followed after NLS1 phase.
The period of BLS1 is estimated  to be $10^{7-8}$ yr
(e.g., Martini 2004; Jakobsen et al. 2003; Croom et al. 2005),
which is still rather shorter than a few Gyr leaving
the problem of bar destruction unsolved.
To summarize, the time-scale for the bar destruction is expected to be
rather longer than  that of the NLS1 and BLS1 phases.
Thus, we may need a new idea to destroy the bar structure
rapidly, though it may not be serious because we do not need to dissolve 
all the bars of NLS1s.

\section{SUMMARY}
Based on multi-wavelength observations and theoretical modeling,
NLS1s are likely in an early phase
of a super-massive BH evolution.
Less massive BHs with high accretion rates suggest
that BHs in NLS1s have not yet been fed enough to
become massive ones, and their BHs are now rapidly
growing (e.g., Mathur 2000; Wandel 2002; Kawaguchi et al. 2004).
If NLS1s are indeed in the early phase of BH evolution,
they can be key objects for studying formation and
evolution of AGNs (Wandel 2002).
Revealing morphology of their host galaxies in their
early phase of AGN activity would be important to 
examine trigger mechanism(s) of AGNs.
Motivated by the idea, we started imaging observations of
nearby 50 NLS1s, which were all the known NLS1s at $z \leq0.0666$ 
(and $\delta \ge -25^{\circ}$) at the time of
starting this program in 2001.
We obtained 40 new images mainly with UH88 inch telescope in
$I_{\rm C}$-band and with CFHT in $g'$-band.
Combining additional 10 images from archive data,
we presented the optical images of these 50 NLS1s.

With these imaging data, we made morphology classification
of them by eye inspection and by quantitative classification
(radial variation of ellipticity and position angle of isophotes).
Based on the classifications, we derived frequency of global
bar structure among the NLS1s. 
It is found that their host galaxies have the bar structure 
in the optical bands more frequently ($85 \pm 7$\%) 
among disk galaxies than BLS1s  ($60-70$\%) and  normal disk 
galaxies ($40-70$\%),  confirming the results by 
Crenshaw et al. (2003) with a similar analysis for 19 NLS1s.
The significance is, however,  marginal particularly when we consider
the uncertainty of the classification.
The bar frequency is comparable to that of HII/starburst galaxies 
($\sim 80$\%).
We also examine the bar frequencies of NLS1s against 
FWHM of the H$\beta$ broad emission-line, Eddington ratio, and
BH mass.
Although  possible correlations that the bar frequency in
the total sample increases with decreasing FWHM and  
with increasing Eddington ratio are seen in the eye inspection 
classification, 
the trends are not seen in quantitative classification,
and the no trend is consistent with the resulting bar frequencies
in the eye inspection classification.
An outer ring structure in NLS1s is very rare in this study as
compared with the BLS1s.
However the completeness of the observations may not be good
enough to detect outer ring  and a further study is 
desirable to be conclusive.

If the difference of  morphology frequency of the host galaxies 
of NLS1s against BLS1s (more frequent bars and rare outer rings)
is significant,  it argues against the idea that a NLS1 is a BLS1 with a
different viewing angle. 
If the difference of the bar frequency against BLS1s and
the similarity of it to HII/starburst galaxies
(comparable bar frequency) are significant, these suggest that 
a NLS1 phase starts from
a starburst phase and evolves to a BLS1 phase after the NLS1 phase.
Further, a plausible mechanism for AGN trigger can either a galactic
bar structure or something else that has excited a bar,
and the bar would fuel the central BH and form the outer ring
by transforming angular momentum of gas.
The bar destruction seems to be possible in terms of 
the necessary mass concentration estimated by numerical simulations.
However, the time-scale necessary to dissolve the bar structure seems to
be rather  longer than the periods of NLS1 and BLS1 phases,
which might challenge to current understanding of destruction 
process(es) of the  bar.

\acknowledgments\
We are grateful to the staff members of UH 88 inch telescope.
Use of the UH 88 inch telescope for the observations
is supported by National Astronomical Observatory of Japan (NAOJ).
We are also grateful to the staff members of CFHT, Subaru,
and KPNO.
We appreciate the support by Masayuki Akiyama 
for taking and reducing a part of the data,
and by Ohad Shemmer and the stuff members of the Wise Observatory 
in the early beginning of this program (Nov., 2000 -- Sep. 2001).
We thank John Tonry for providing us the conflat program
to reduce the OPTIC data and Masafumi Noguchi for useful discussion.
We also thank referee's comments which improve this paper very much.
This research  made use of the NASA/IPAC Extragalactic Database,
which is operated by the Jet Propulsion Laboratory, California Institute
of Technology, under contract with the National Aeronautics and Space
Administration.
K.O. is supported by a Grant-in-Aid for Scientific Research from 
Japan Society for Promotion of Science (17540216).
T.K. thanks the financial supports from Postdoctoral Fellowships of 
the Japan Society for the Promotion of Science
and from The Research Institute of Aoyama Gakuin University.

\appendix
\section{RX J0024.7+0820 at $z=0.067$}
During the course of the imaging program, we took an image of
RX J0024.7+0820 with FOCAS on Subaru, though the redshift (0.067) is 
slightly larger than our criteria (0.0666) (Xu et al. 1999).
Figure \ref{rxj0024} shows a clear bar structure with outer-ring-like
arms.
The redshift determined by ourselves based on the KPNO spectrum is
0.0671.

\clearpage

\begin{deluxetable}{lcccccccccc}
\tabletypesize{\scriptsize}
\rotate
\tablecaption{Sample Narrow-Line Seyfert 1 Galaxies \label{tblsample}}
\tablewidth{0pt}
\tablehead{
\colhead{Name} & \colhead{FWHM H$\beta$} & \colhead{Reference} &  
\colhead{$z$} & \colhead{$B^a$} &
\colhead{log $\lambda L_{5100}$} & \colhead{log $M_{\rm BH}$} & 
\colhead{log {$L_{\rm bol} \over L_{\rm Edd}$}} &
\colhead{Morphology$^b$} & \colhead{Bar$^c$} &
\colhead{Data Source$^d$} \\

\colhead{} & \colhead{(km s$^{-1}$)}  & \colhead{} & 
\colhead{} & \colhead{(mag)} &
\colhead{(ergs s$^{-1}$)} & \colhead{($M_{\odot}$)} & 
\colhead{} & \colhead{} & \colhead{} & \colhead{} 
}

\startdata
Zw493.004 &       500      & 7 & 0.043 &15.008 & 44.05  &   6.11   & +0.94& SBb$^e$ & SAB & POSS \\
Mrk 493  &        740      & 1 & 0.031 &15.584 & 43.53  &   6.09   & +0.44& SBb (SB/S(B)a) & SB & UH88-1 \\
IRAS 05262+4432 & 740      & 1 & 0.032 &11.412 & 45.22  &   7.28   & +0.95& SBb & SB & UH88-2 \\
PG 1244+026 &     740      & 1 & 0.048 &16.036 & 43.73  &   6.23   & +0.50& E/S0 & no & CFHT  \\
RX J1618.1+3619 & 830      & 4 & 0.034 &16.547 & 43.22  &   5.97   & +0.25& S0,dE? & no & UH88-1 \\
IRAS 04312+4008 & 860      & 1 & 0.020 &11.887 & 44.62  &   6.98   & +0.64& SBb & SB & UH88-2 \\
Mrk 42   &        865      & 1 & 0.024 &16.114 & 43.09  &   5.92   & +0.17& SBab (SB/SBa) & SB & CFHT \\ 
Akn 564  &        865      & 1 & 0.025 &14.602 & 43.73  &   6.37   & +0.37& SBa & SB & UH88-2 \\   
Mrk 359  &        900      & 1 & 0.017 &14.656 & 43.37  &   6.15   & +0.22& SBa & SB & UH88-2 \\
KUG 1031+398 &    935      & 1 & 0.042 &15.534 & 43.82  &   6.49   & +0.32& SBa & SB & UH88-2 \\    
B3 1702+457 &     975      & 1 & 0.061 &14.951 & 44.38  &   6.92   & +0.46& SBab & SB & UH88-1 \\

Mrk 1044 &       1010      & 1 & 0.016 &14.593 & 43.34  &   6.23   & +0.11& SAB0/a (S/Sa) & SB & UH88-2 \\
RX J1531.6+2019 & 1050     & 7 & 0.051 &16.907 & 43.44  &   6.33   & +0.11& E/S0 & no & UH88-3 \\
PG 1448+273 &    1050      & 1 & 0.065 &14.885 & 44.46  &   7.05   & +0.42& Int & no & CFHT \\   
Mrk 1239  &      1075      & 1 & 0.019 &14.708 & 43.45  &   6.36   & +0.09& S0 & no & CFHT \\
TON S180  &      1085      & 1 & 0.062 &14.538 & 44.56  &   7.14   & +0.42& SABa & SB & UH88-2 \\
IZw1   &         1090      & 7 & 0.061 &14.131 & 44.71  &   7.25   & +0.46& SABb (SB/-) & no & HST \\
RX J0032.3+2423 & 1110     & 3 & 0.066 &17.158 & 43.57  &   6.47   & +0.10& S0 & SB &  UH88-2 \\
NGC 4051 &       1120      & 1 & 0.002 &13.534 & 41.95  &   5.35   & $-$0.39& SBb (SB/Sb) & $-$ & CFHT \\
Mrk 896  &       1135      & 1 & 0.027 &15.074 & 43.61  &   6.52   & +0.09& SABc (S/Sc) & SB & HST \\
MCG 06.26.012 &  1145      & 1 & 0.033 &15.318 & 43.69  &   6.58   & +0.11& SBb (SB/SB0) & SB & CFHT \\
Mrk 684  &       1150      & 1 & 0.046 &15.271 & 44.00  &   6.80   & +0.20& SBab & SB & CFHT \\
IRAS 04576+0912 & 1210     & 1 & 0.037 &15.89  & 43.56  &   6.54   & +0.02& SBa & SB & UH88-2 \\
PG 0934+013(Mrk707) & 1295 & 1 & 0.051 &16.299 & 43.68  &   6.68   & $-$0.00& SBab & SB & UH88-2 \\
RX J17450+4802 & 1355      & 1 & 0.054 &16.192 & 43.78  &   6.79   & $-$0.01& SABb? & no & POSS  \\
Mrk 142  &       1370      & 1 & 0.045 &16.141 & 43.63  &   6.70   & $-$0.06& SB0/a & SB & UH88-2 \\ 
PG 1011-040 &    1455      & 1 & 0.058 &15.331 & 44.18  &   7.13   & +0.05&  SBb$^e$ & SB & POSS \\
IRAS 15091-2107 & 1460     & 6 & 0.044 &15.223 & 43.98  &   7.00   & $-$0.01& E/S0 & no & UH88-3 \\
RX J1032.7+3913 & 1460     & 3 & 0.064 &16.534 & 43.79  &   6.86   & $-$0.07& E/S0?  & no & SDSS DR5 \\
KAZ 320  &       1470      & 1 & 0.034 &16.128 & 43.39  &   6.59   & $-$0.20& E/S0 & no & UH88-2 \\
  
RX J1402.5+2159 & 1520     & 3 & 0.066 &16.602 & 43.79  &   6.90   & $-$0.11& E/S0? & no & SDSS DR5 \\
RX J0140.1+1129 & 1530     & 3 & 0.065 &15.923 & 44.05  &   7.08   & $-$0.03& P & $-$ & Subaru \\
HS 1831+5338 &   1555      & 1 & 0.039 &15.746 & 43.66  &   6.83   & $-$0.16& SB0 & SB & UH88-1 \\
NGC 4748 &       1565      & 1 & 0.014 &14.507 & 43.26  &   6.55   & $-$0.29& SABb,Int (S/Sa) & SB & CHFT \\
Mrk 507  &       1565      & 1 & 0.053 &16.162 & 43.77  &   6.91   & $-$0.14& Int? & $-$ & UH88-1 \\
RX J1016.7+4210 & 1570     & 3 & 0.056 &16.533 & 43.67  &   6.84   & $-$0.17& S0/a,int? & SB & UH88-2 \\
PG 1016+336 &    1590      & 1 & 0.024 &15.812 & 43.21  &   6.53   & $-$0.32& SBb & SB & CFHT \\
NPM1G -14.0512 & 1605      & 1 & 0.042 &14.753 & 44.13  &   7.18   & $-$0.05& SBb & SAB & CFHT \\
Mrk 739E  &      1615      & 1 & 0.030 &14.752 & 43.83  &   6.98   & $-$0.15& Int & $-$ & CFHT \\
Mrk 766  &       1630      & 1 & 0.012 &14.256 & 43.23  &   6.56   & $-$0.34& SBb (SB/SBc) & SB & HST \\
PG 1535+547(Mrk486) & 1680 & 1 & 0.038 &15.147 & 43.88  &   7.05   & $-$0.17& SAB0? & SB & CFHT \\
RX J0000.1+0523 & 1690     & 3 & 0.040 &16.163 & 43.52  &   6.80   & $-$0.28& SB0? & SB & UH88-2 \\
KUG 1618+410   &  1700     & 7 & 0.038 &15.969 & 43.55  &   6.83   & $-$0.28& SBab & SB & SDSS DR5 \\
CTS J03.19 &  1735$^{f}$   & 2 & 0.053 &15.463 & 44.05  &   7.19   & $-$0.14& SBab & SB & UH88-2  \\
PG 0923+129(Mrk705) & 1790 & 1 & 0.028 &14.974 & 43.68  &   6.96   & $-$0.28& RS0 & no & CFHT \\
PG 1119+120(Mrk734) & 1825 & 1 & 0.049 &15.241 & 44.07  &   7.25   & $-$0.18& SABa & SB & CFHT \\
Mrk 335 &        1851      & 5 & 0.025 &14.037 & 43.96  &   7.18   & $-$0.23& P (P/?) & no & HST \\
KAZ 163 &        1875      & 1 & 0.063 &15.308 & 44.27  &   7.41   & $-$0.15& Int? & $-$ & UH88-1 \\
WAS 61 &        1900       & 4 & 0.045 &15.317 & 43.96  &   7.21   & $-$0.25& S0? & SB & SDSS DR5 \\
CG 59  &         1990      & 4 & 0.049 &15.57  & 43.94  &   7.23   & $-$0.30& SB0a & SB & UH88-2 \\

\enddata

\tablenotetext{a}{$B$ magnitudes taken from V\'{e}ron-Cetty \& V\'{e}ron
(2003) and corrected for the Galactic extinction by using NED (Schlegel et al. (1998)).}
\tablenotetext{b}{Morphology in parentheses is that assigned by 
Crenshaw et al. (2003) and Malkan et al. (1998) by separating with `/'.}
\tablenotetext{c}{Quantitative classification for the presence of a bar 
(see text  for detail). 
SB, non, and $-$ refer to the presence of a bar, the absence of a
 bar, and no answer (due to interacting feature or the
 presence of strong asymmetry),  respectively.
SAB denotes an object which does not satisfy the bar criteria but
 almost satisfy the criteria (see text for details). }
\tablenotetext{d}{See also Table\ref{tblrun}.}
\tablenotetext{e}{Taken from NED after checking POSS image by ourselves.}
\tablenotetext{f}{FWHM of the H$\alpha$ broad-emission line.}
\tablerefs{
(1) V\'{e}ron-Cetty et al. 2001; 
(2) Rodr\'iguez-Ardila et al.  2000;
(3) Xu et al. 1999; 
(4) Grupe et al. 1999; 
(5) Marziani et al. 2003;
(6) Boller et al. 1996; 
(7) our data taken with GoldCam at KPNO 2.1m.
}
\end{deluxetable}

\clearpage
\begin{deluxetable}{cccccc}
\tabletypesize{\scriptsize}
\tablecaption{Journal of Observations\label{tblrun}}
\tablewidth{0pt}
\tablehead{
\colhead{Observing Run} & \colhead{Date} & 
\colhead{Instrument} & \colhead{Pixel Size} &
\colhead{Band} & \colhead{Seeing} \\

\colhead{} & \colhead{}   & 
\colhead{} & \colhead{(arcsec pixel$^{-1}$)} &
\colhead{} & \colhead{(arcsec)} 
}

\startdata
UH88-1 & Apr. 2003 & OPTIC   & 0.276 & $I_c$ & 0.7--1.0 \\
UH88-2 & Dec. 2004 & OPTIC   & 0.138 & $I_c$ & 0.5--1.0 \\
UH88-3 & May  2005 & Tek2k   & 0.220 & $I_c$ & 0.7--1.2 \\
CFHT   & Mar. 2003 & MegaCam & 0.185 & g$^{\prime}$ & 0.8--1.4 \\
Subaru & Nov. 2004 & FOCAS   & 0.104 & $I_c$ & 0.6--1.1 \\
\enddata
\end{deluxetable}

\clearpage

\begin{deluxetable}{cccc}
\tabletypesize{\scriptsize}
\tablewidth{0pt}
\tablecolumns{3}
\tablecaption{Bar (SB+SAB) Frequencies for Sub-Samples based on the
Classification by Eye Inspection\label{tblfreq}}
\tablehead{
\colhead{Sample} & 
\colhead{Bar frequency [\%] (numbers)} &
\colhead{Bar frequency [\%] (numbers)}\\
\colhead{} & 
\colhead{Total sample} &
\colhead{Disk sample}}
\startdata
all bars &  $64 \pm 7$ (32/50) & $84 \pm 6$ (32/38) \nl
\hline
 \qquad\quad\, FWHM(H$\beta$) $<$ 1000 km s$^{-1}$ & $82 \pm 12$ (9/11) & $90 \pm 9$ (9/10) \nl
 1000 $\leq$ FWHM(H$\beta$) $<$ 1500 km s$^{-1}$  & $63 \pm 11$ (12/19)& $86 \pm 9$ (12/14) \nl 
 1500 $\leq$ FWHM(H$\beta$) $<$ 2000 km s$^{-1}$  & $55 \pm 11$ (11/20)& $79 \pm 11$ (11/14) \nl 
\hline
 $-0.39 \leq$ log $L_{\rm bol}$/$L_{\rm Edd}$ $< -0.1$  & $55 \pm 11$ (11/20) & $79 \pm 11$ (11/14) \nl
 $-0.1 \leq$  log $L_{\rm bol}$/$L_{\rm Edd}$ $< -0.3$  & $63 \pm 11$ (12/19) & $80 \pm 10$ (12/15) \nl
 $0.3 \leq$ log $L_{\rm bol}$/$L_{\rm Edd}$   $< 1.0$     & $82 \pm 12$ (9/11) & $100$ (9/9) \nl
\hline
 5.35 $\leq$ log $M_{\rm BH} (M_{\odot})$ $< 6.5$     & $62 \pm 13$ (8/13) & $73 \pm 13$ (8/11) \nl
 6.5  $\leq$ log $M_{\rm BH} (M_{\odot})$ $< 7.0$     & $68 \pm 10$ (15/22) & $88 \pm 8$ (15/17) \nl
 7.0  $\leq$ log $M_{\rm BH} (M_{\odot})$ $< 7.5$     & $60 \pm 13$ (9/15) & $90 \pm 9$ (9/10) \nl

\enddata
\end{deluxetable}

\clearpage

\begin{deluxetable}{cccc}
\tabletypesize{\scriptsize}
\tablewidth{0pt}
\tablecolumns{3}
\tablecaption{Bar (SB+SAB) Frequencies for Sub-Samples based on 
the quantitative Classification\label{tblfreq2}}
\tablehead{
\colhead{Sample} & 
\colhead{Bar frequency [\%] (numbers)} &
\colhead{Bar frequency [\%] (numbers)}\\
\colhead{} & 
\colhead{Total sample} &
\colhead{Disk sample}}
\startdata
all bars & $71 \pm 7$ (32/45) & $86\pm6$ (32/37) \nl
\hline
 \qquad\quad\,  FWHM(H$\beta$)  $<$ 1000 km s$^{-1}$ & $82 \pm 12$ (9/11) & $90 \pm 9$ (9/10) \nl
 1000 $\leq$ FWHM(H$\beta$) $<$ 1500 km s$^{-1}$  & $56 \pm 12$ (10/18)& $77 \pm 12$ (10/13) \nl 
 1500 $\leq$ FWHM(H$\beta$) $<$ 2000 km s$^{-1}$  & $81 \pm 10$ (13/16)& $93 \pm 7$ (13/14) \nl 
\hline
 $-0.39 \leq$ log $L_{\rm bol}$/$L_{\rm Edd}$ $< -0.1$  & $75 \pm 11$ (12/16) & $92 \pm 7$ (12/13) \nl
 $-0.1  \leq$ log $L_{\rm bol}$/$L_{\rm Edd}$ $<  0.3$    & $67 \pm 11$ (12/18) & $80 \pm 10$ (12/15) \nl
 $0.3   \leq$ log $L_{\rm bol}$/$L_{\rm Edd}$ $<  1.0$     & $73 \pm 13$ (8/11) & $89 \pm 10$ (8/9) \nl
\hline
 5.35 $\leq$ log $M_{\rm BH} (M_{\odot})$ $<$ 6.5     & $67 \pm 14$ (8/12) & $80 \pm 13$ (8/10) \nl
 6.5  $\leq$ log $M_{\rm BH} (M_{\odot})$ $<$ 7.0      & $75 \pm 10$ (15/20) & $88 \pm 8$ (15/17) \nl
 7.0  $\leq$ log $M_{\rm BH} (M_{\odot})$ $<$ 7.5      & $69 \pm 13$ (9/13) & $90 \pm 9$ (9/10) \nl

\enddata
\end{deluxetable}

\clearpage

\begin{figure}[p]
\caption{
Optical images of all (50) narrow-line Seyfert 1s in our sample 
which covers all the known NLS1s at $z<0.0666$ (and $\delta \ge
-25^{\circ}$) at the time of 2001. The images are in the order
of H$\beta$ emission-line width from smaller FWHM to larger FWHM 
(same as the order 
in Table 1). In each image, a field of view of 50$^{\prime\prime} 
\times 50^{\prime\prime}$ is shown,which corresponds to 64\,kpc 
$\times$ 64\,kpc at $z=0.0666$.  For NGC 4051, a field of view of 
$240^{\prime\prime} \times 240^{\prime\prime}$ is shown.  North is
at the top and east to the left.  The sources of the images are listed
 in Table 1.  
Right panels for each object show a surface brightness 
distribution (upper panel) and radial variation of ellipticity 
and of position angle (lower panel) derived by ellipse fitting 
to the  isophotes of each galaxy.
Solid circles and solid squares show variation of 
ellipticity with left ordinate and of position 
angle with right ordinate, respectively.  Note that the 
spans of the ordinates are not fixed.  Horizontal dotted line, 
dashed line, and dot-dashed line show $e=0.25$, global maximum ellipticity 
$e_{\rm max}$,
and $e_{\rm max}-0.1$, respectively.
[{\it See the electronic edition of the Journal for a color 
version of this figure.}]}
\end{figure}
\clearpage 
\epsscale{.6}
\plotone{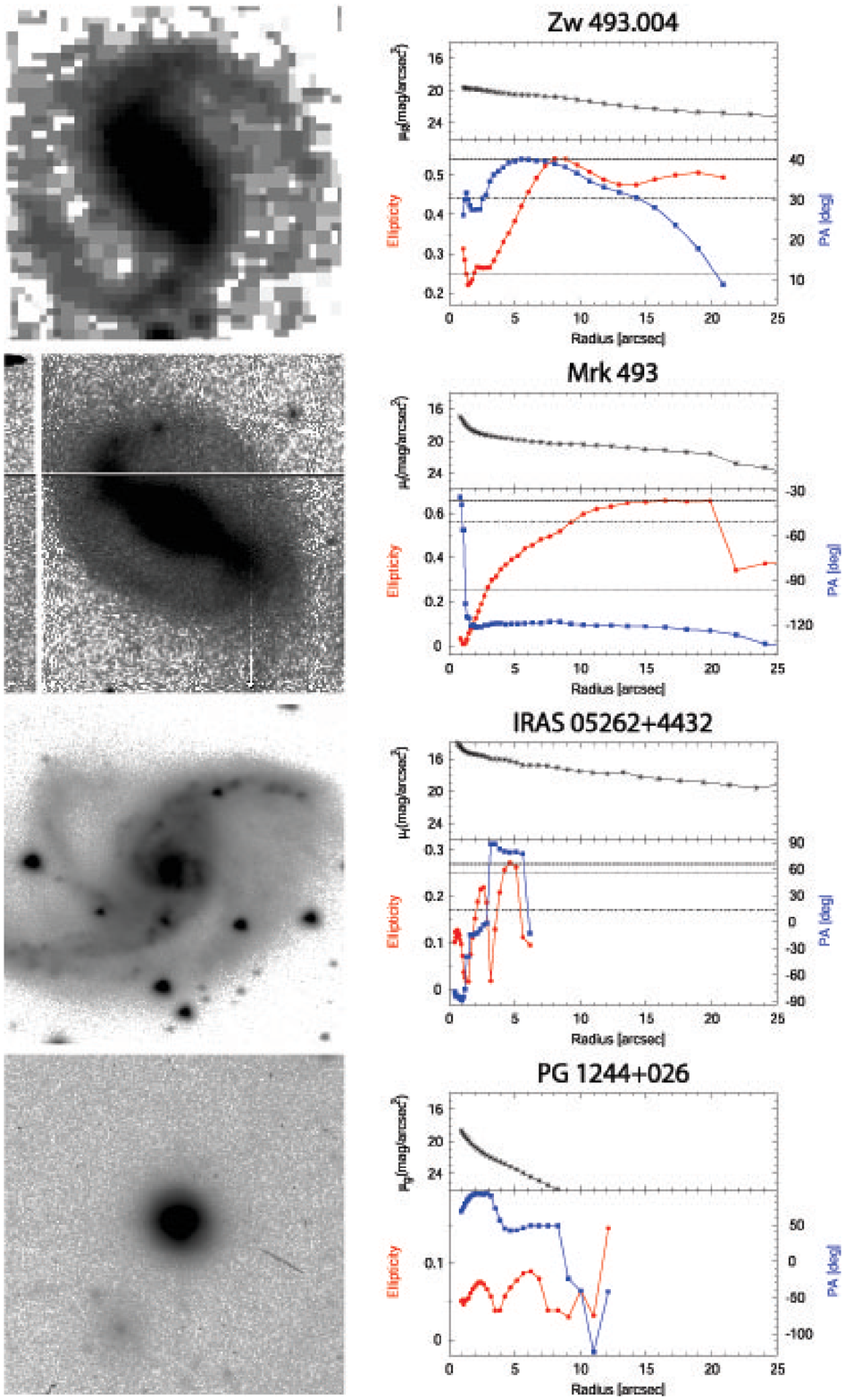}\\[5mm]
\centerline{Fig. 1. ---}
\clearpage
\plotone{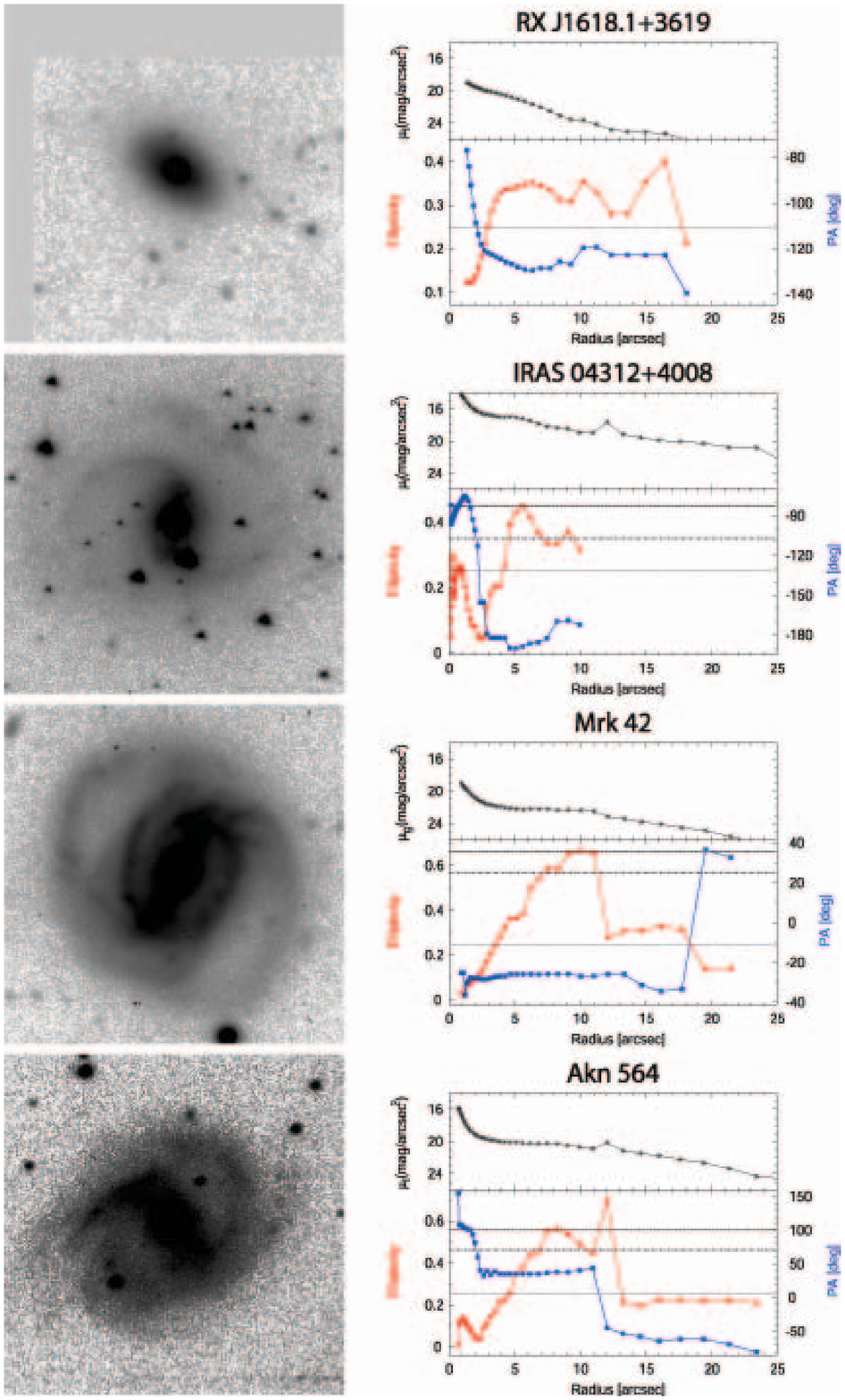}\\[5mm]
\centerline{Fig. 1. ---}
\clearpage
\plotone{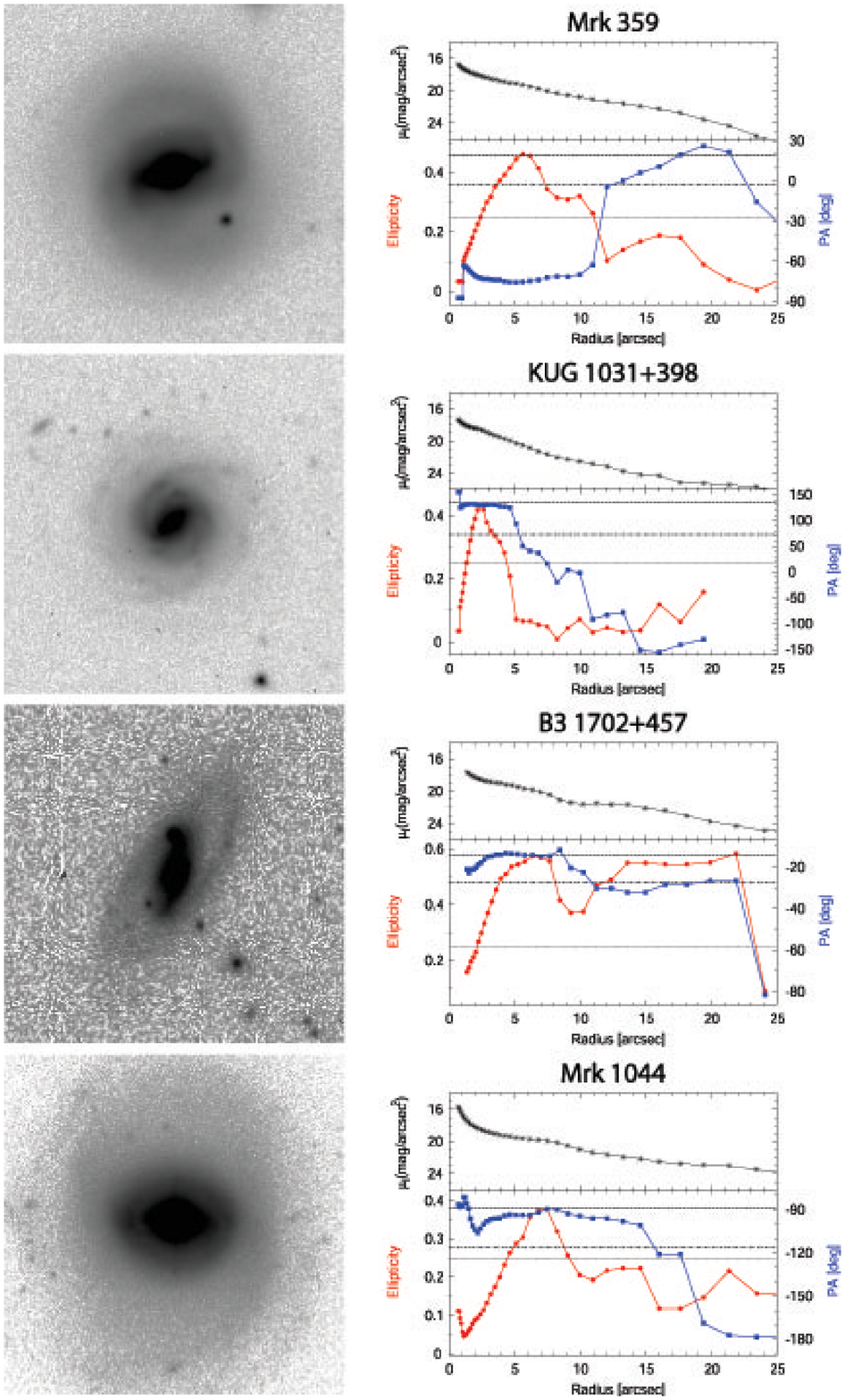}\\[5mm]
\centerline{Fig. 1. ---}
\clearpage
\plotone{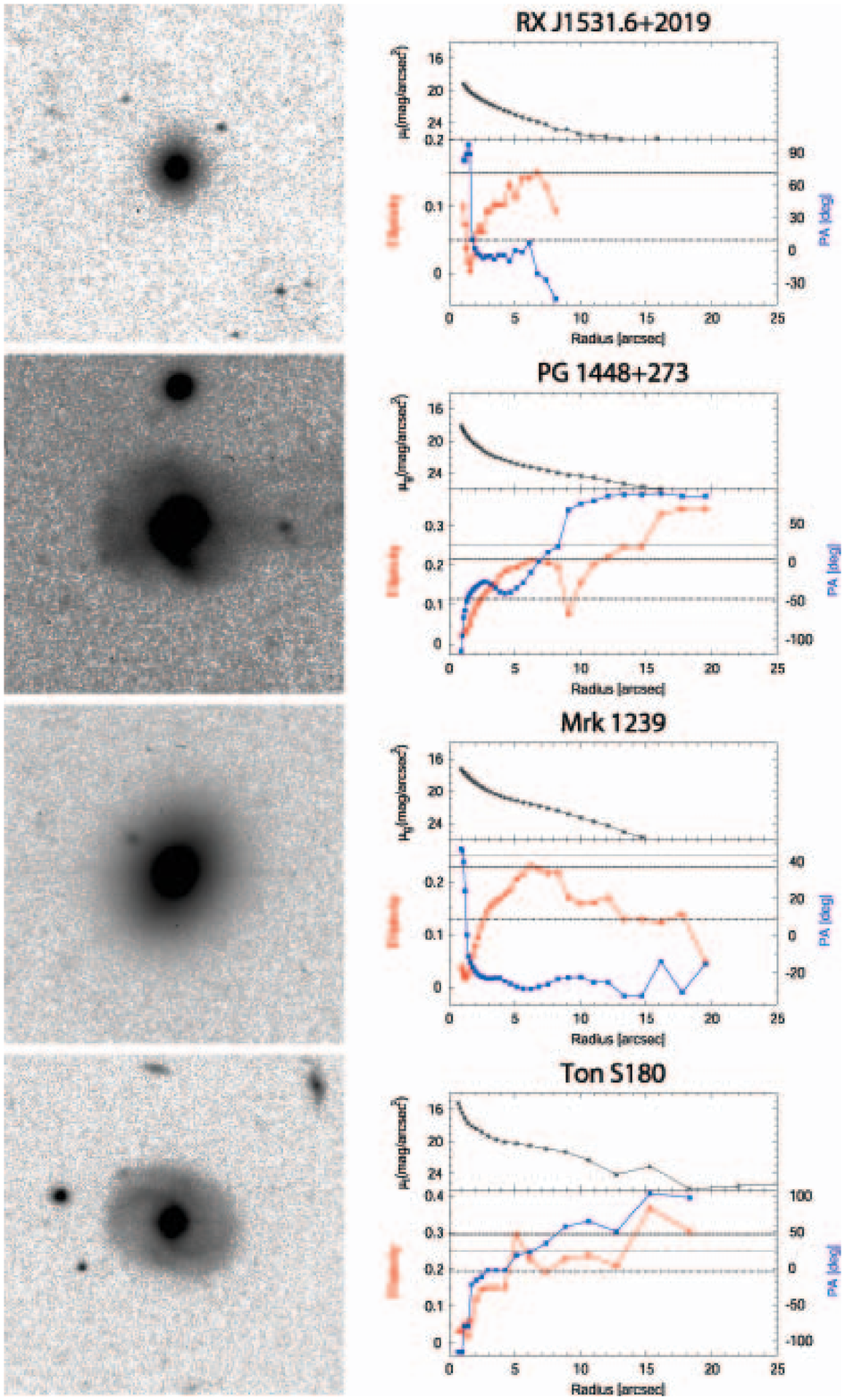}\\[5mm]
\centerline{Fig. 1. ---}
\clearpage
\plotone{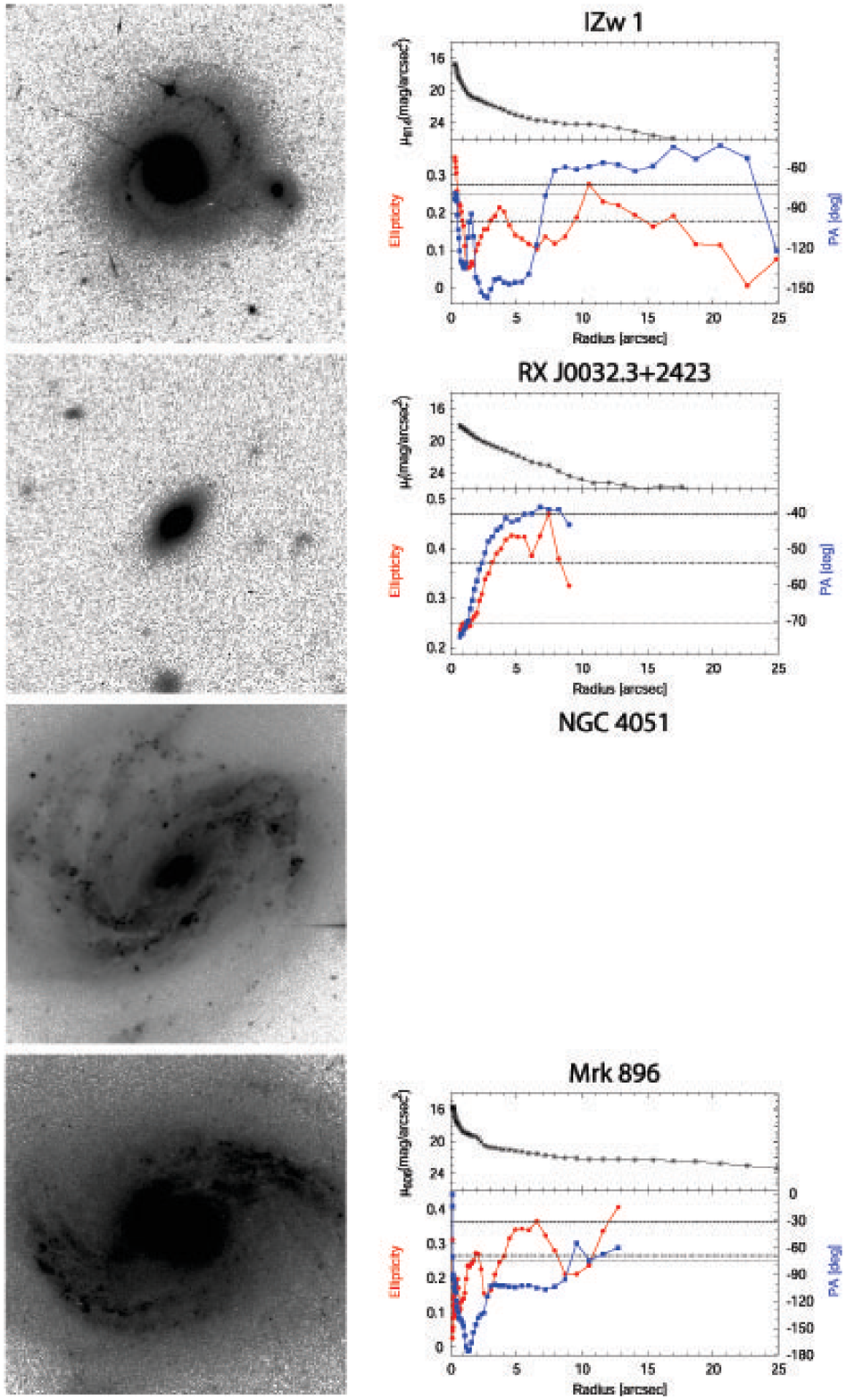}\\[5mm]
\centerline{Fig. 1. ---}
\clearpage
\plotone{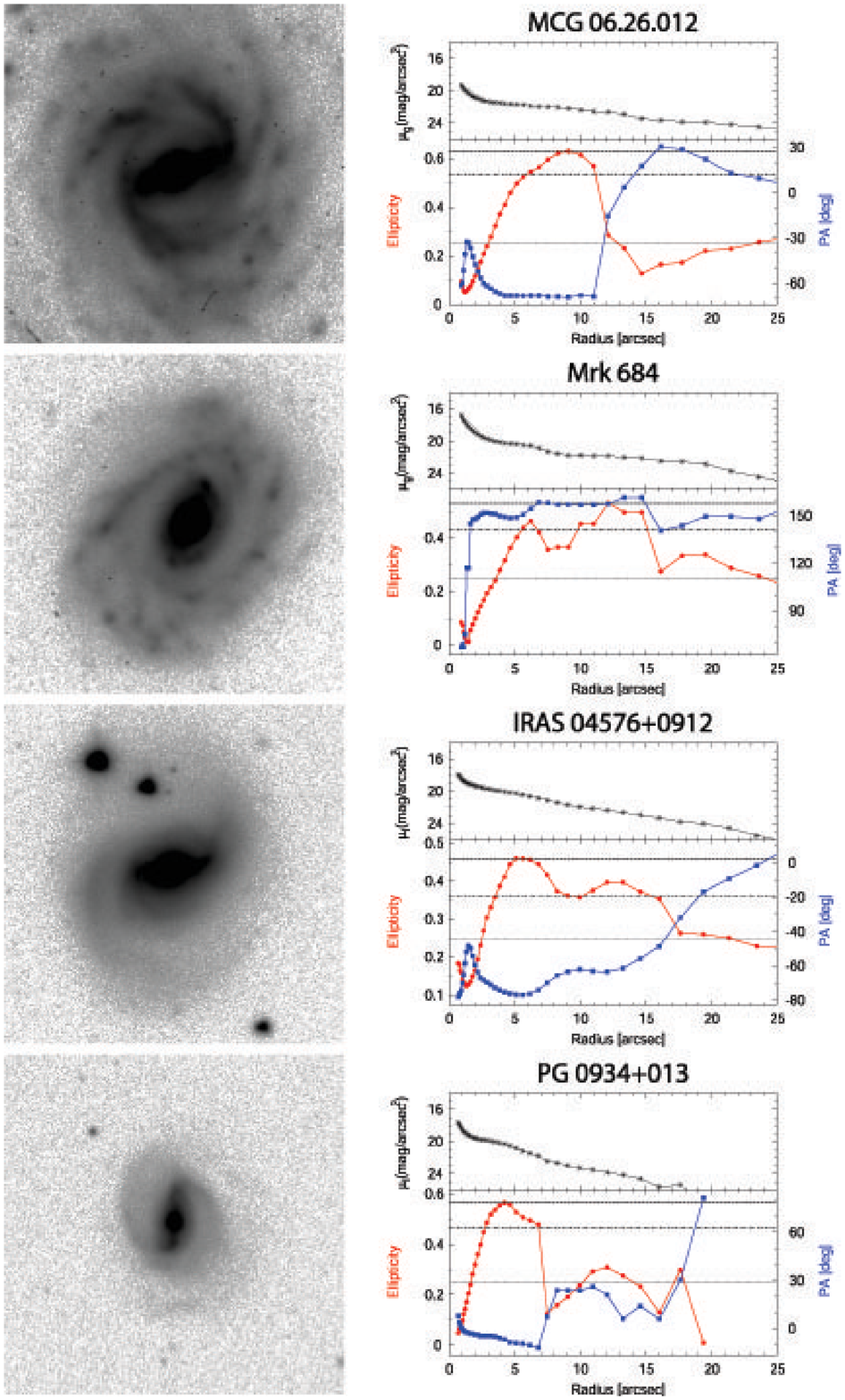}\\[5mm]
\centerline{Fig. 1. ---}
\clearpage
\plotone{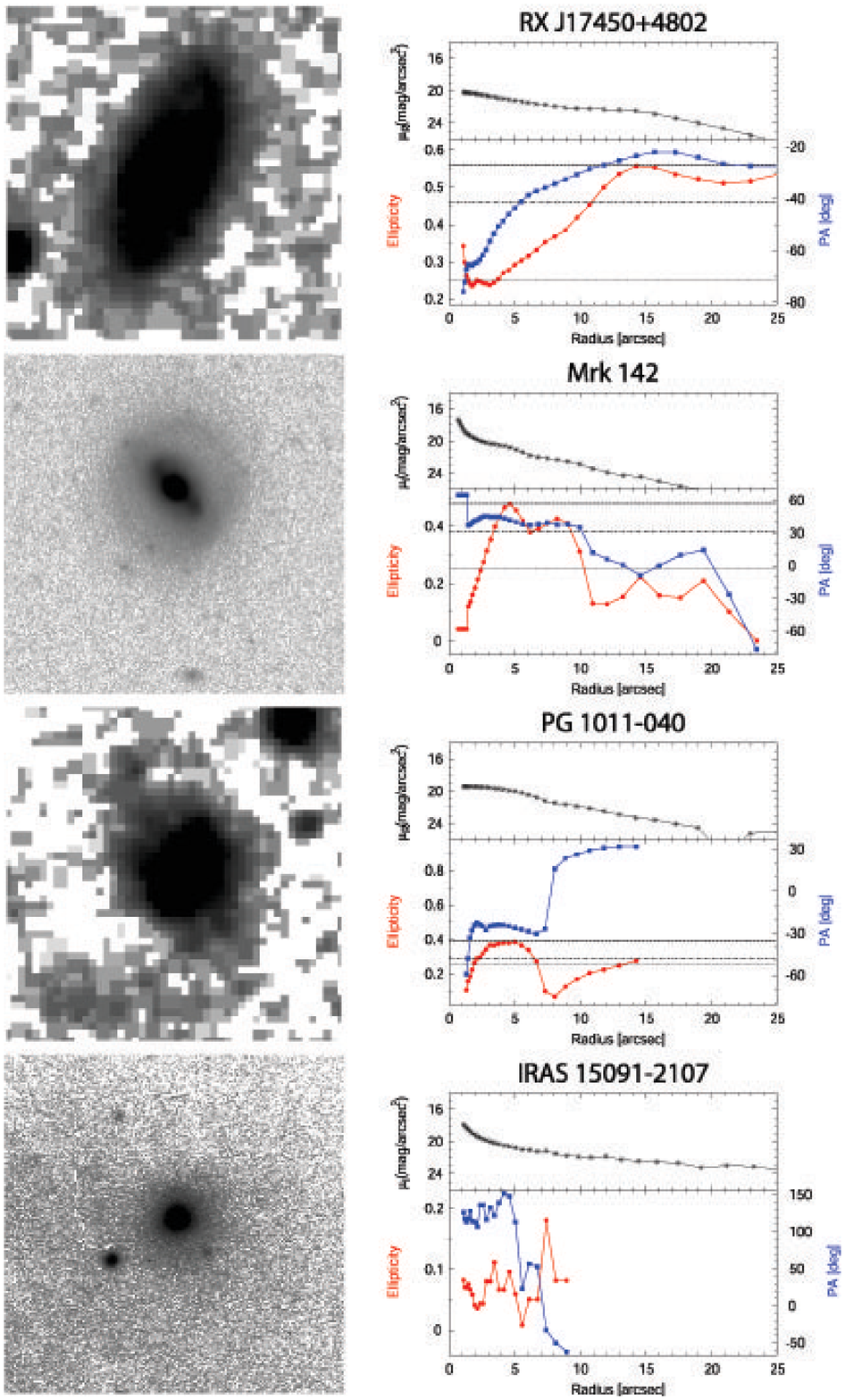}\\[5mm]
\centerline{Fig. 1. ---}
\clearpage
\plotone{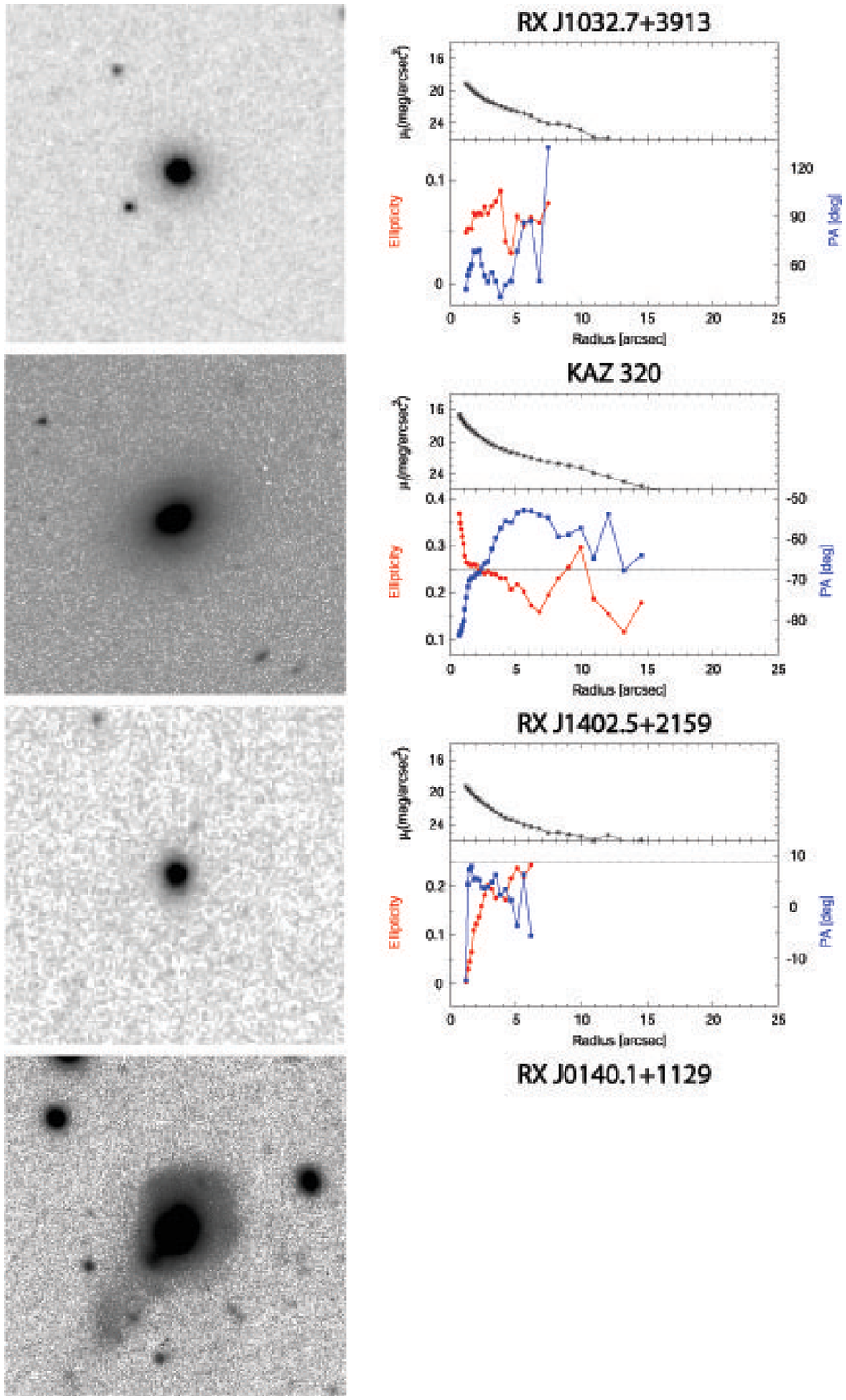}\\[5mm]
\centerline{Fig. 1. ---}
\clearpage
\plotone{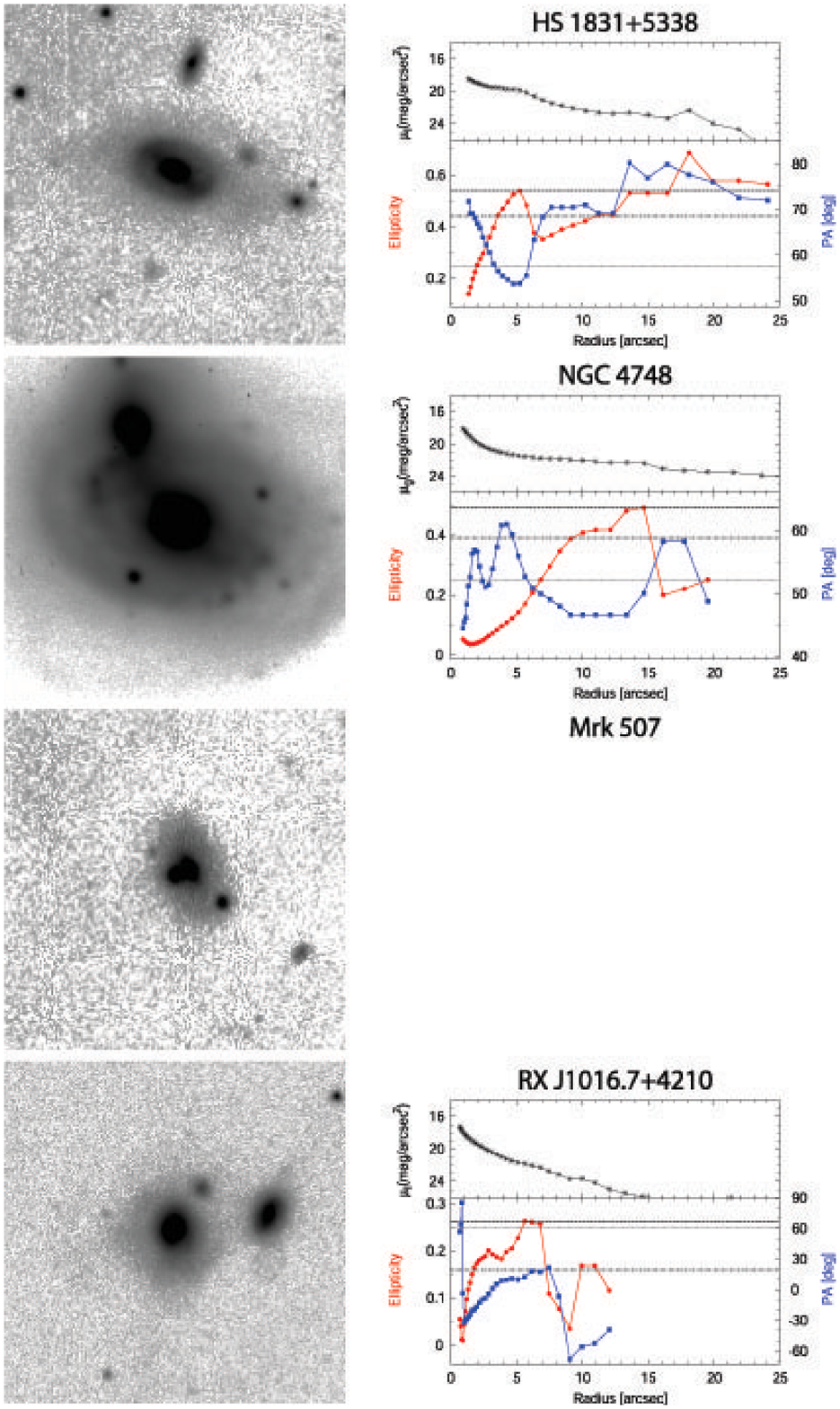}\\[5mm]
\centerline{Fig. 1. ---}
\clearpage
\plotone{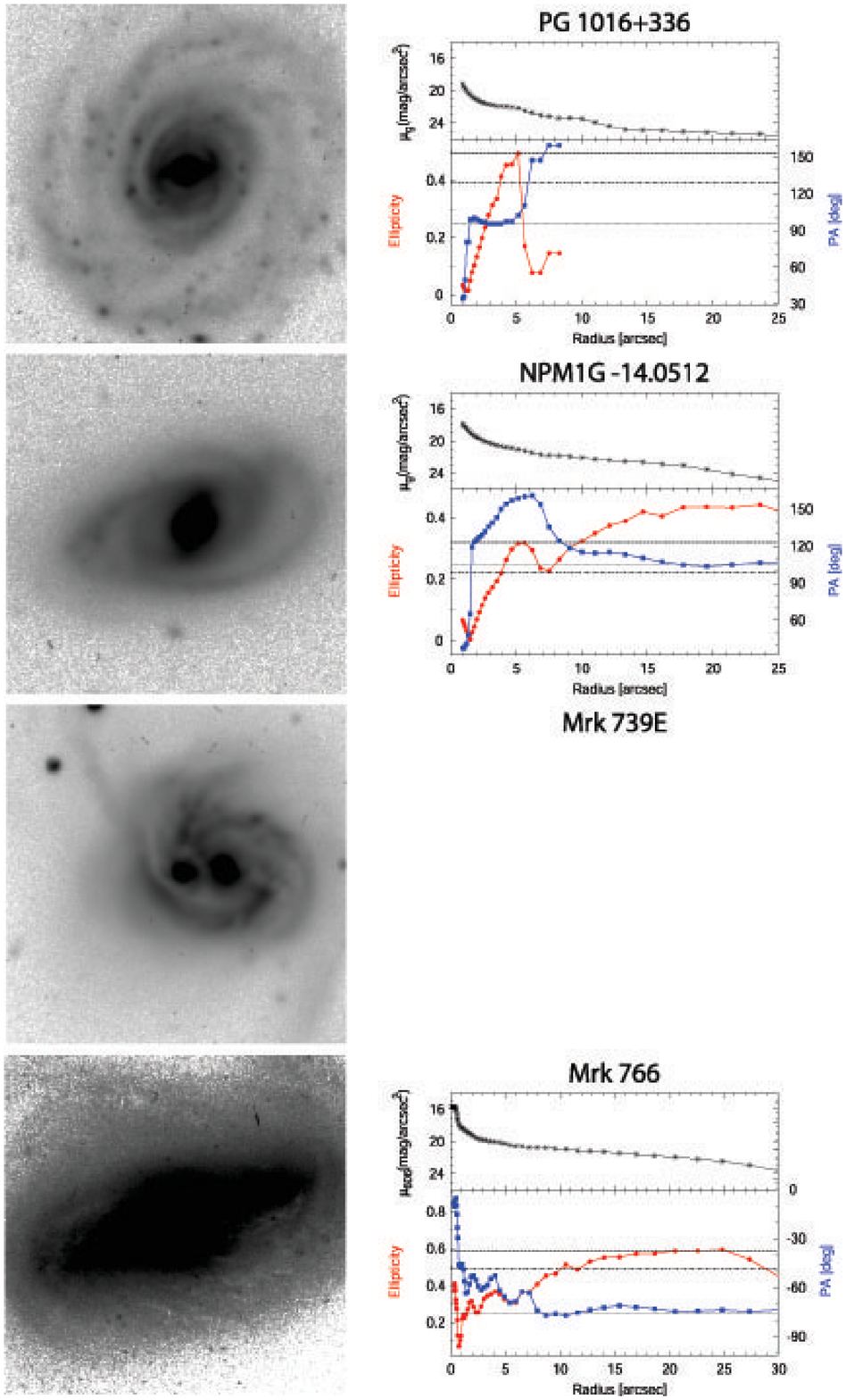}\\[5mm]
\centerline{Fig. 1. ---}
\clearpage
\plotone{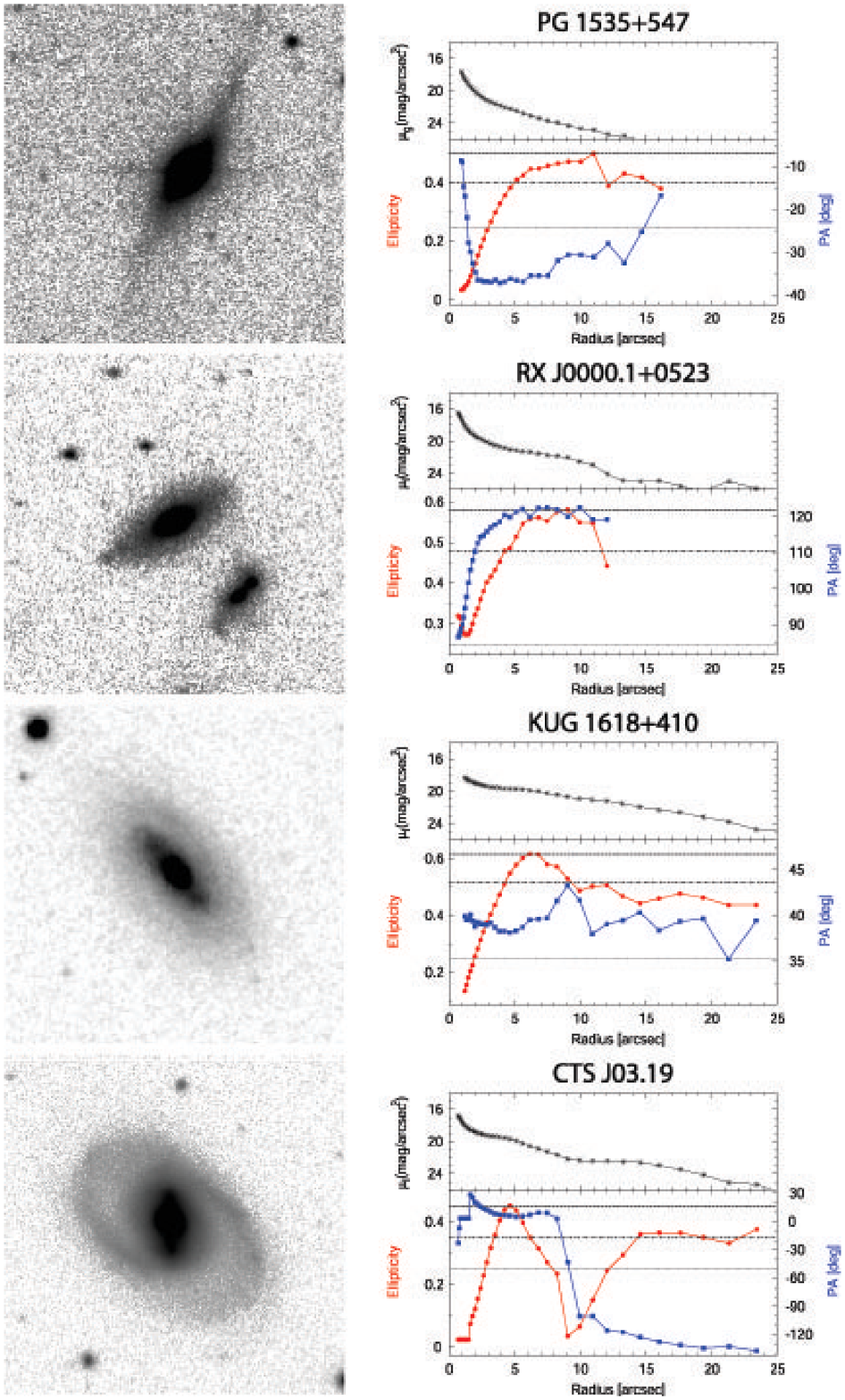}\\[5mm]
\centerline{Fig. 1. ---}
\clearpage
\plotone{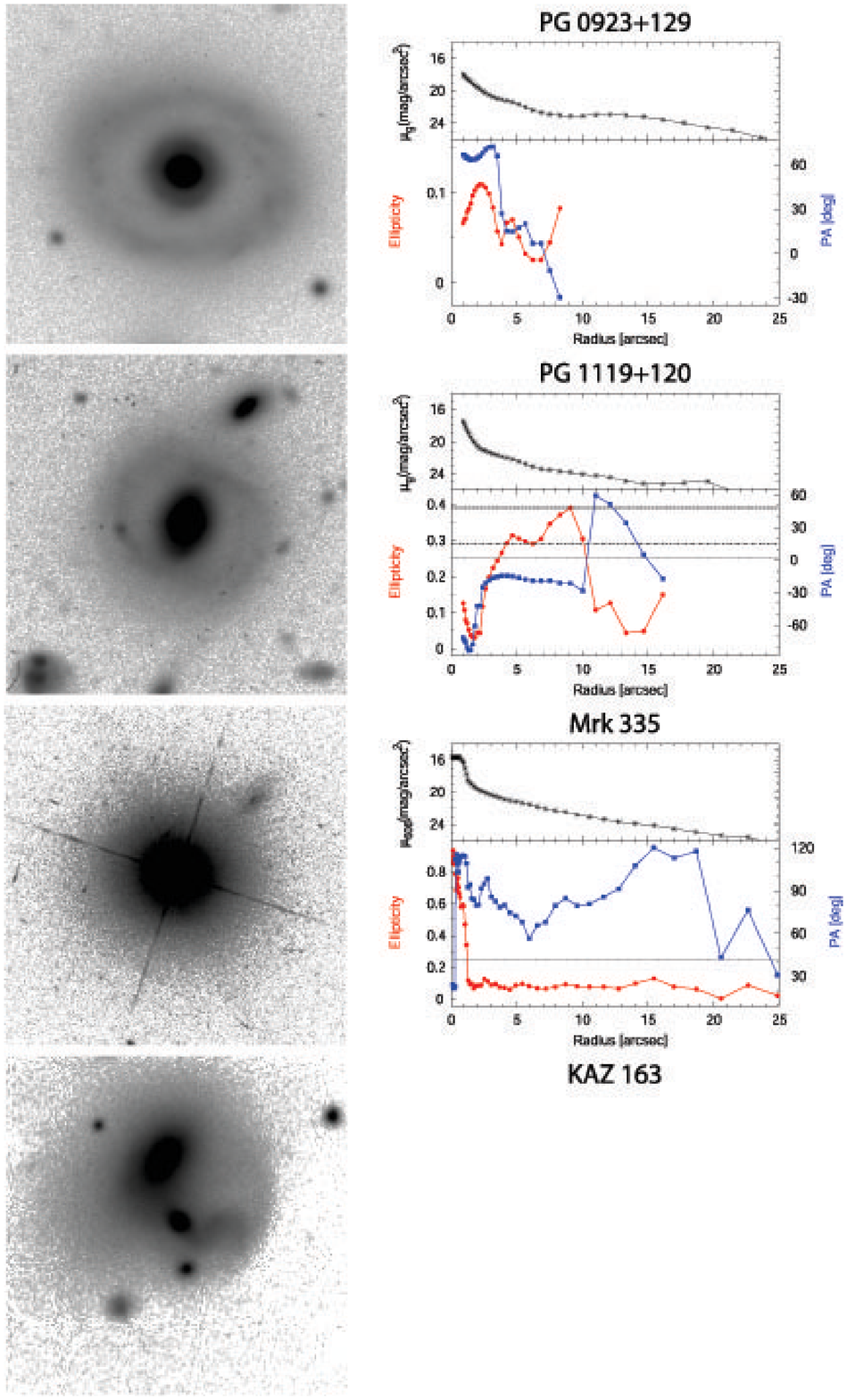}\\[5mm]
\centerline{Fig. 1. ---}
\clearpage
\plotone{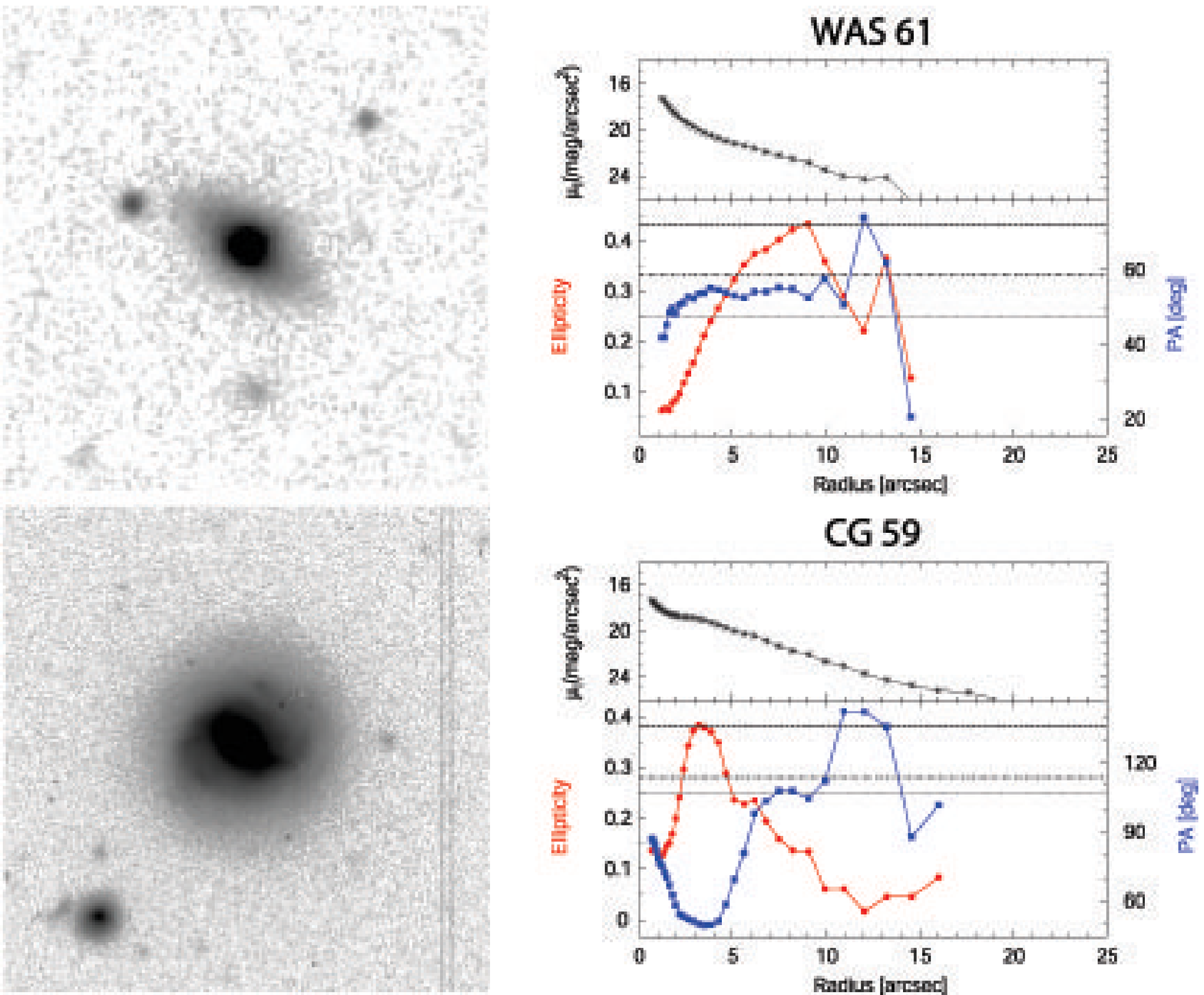}\\[5mm]
\centerline{Fig. 1. ---}
\clearpage

\begin{figure}
\epsscale{1}
\plotone{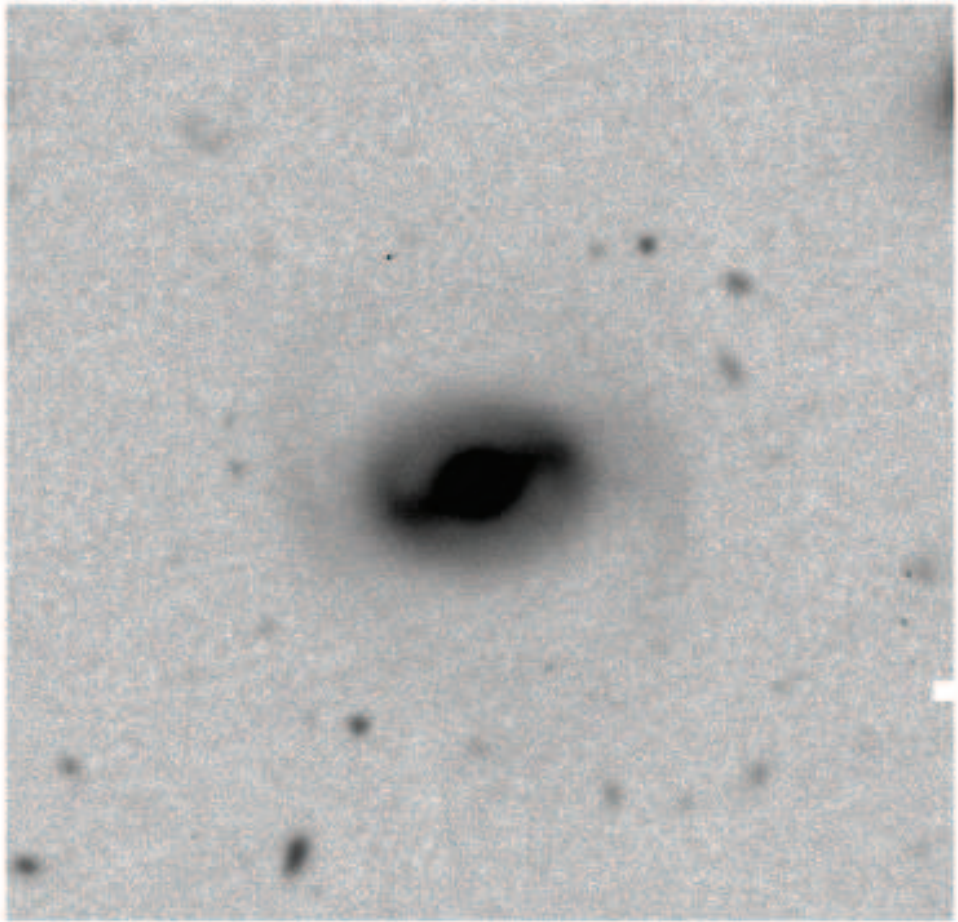}
\caption{
Optical image of RXJ0024.7+0820 ($50^{\prime\prime} 
\times 50^{\prime\prime}$).
North is at the top and east to the left.
\label{rxj0024}}
\end{figure}


\end{document}